\documentclass[prl, twocolumn, superscriptaddress]{revtex4-1}
\usepackage{bm, amsmath, amsfonts, amssymb, braket}
\usepackage{subfigure}
\usepackage{color}
\usepackage{graphicx}
\usepackage{dcolumn} % Align table columns on decimal point

\newcommand{\ii}{\text{i}}

\usepackage{rotating} % for 90 degree rotated table and fig
\def\im{{\rm Im\,}}

\newcommand{\s}{\sigma}
\newcommand{\G}{\Gamma}
\newcommand{\bk}{\bm{k}}
\begin{document}

\title{Topological Origin of Non-Hermitian Skin Effects}

\author{Nobuyuki Okuma}
	\email{okuma@hosi.phys.s.u-tokyo.ac.jp}
	\affiliation{Yukawa Institute for Theoretical Physics, Kyoto University, Kyoto 606-8502, Japan}
\author{Kohei Kawabata}
	\affiliation{Department of Physics, University of Tokyo, 7-3-1 Hongo, Bunkyo-ku, Tokyo 113-0033, Japan}
\author{Ken Shiozaki}
	\affiliation{Yukawa Institute for Theoretical Physics, Kyoto University, Kyoto 606-8502, Japan}
\author{Masatoshi Sato}
	\affiliation{Yukawa Institute for Theoretical Physics, Kyoto University, Kyoto 606-8502, Japan}

\date{\today}

\begin{abstract} 
A unique feature of non-Hermitian systems is the skin effect, which is the extreme sensitivity to the boundary conditions. Here, we reveal that the skin effect originates from intrinsic non-Hermitian topology. Such a topological origin not merely explains the universal feature of the known skin effect, but also leads to new types of the skin effects---symmetry-protected skin effects. In particular, we discover the $\mathbb{Z}_{2}$ skin effect protected by time-reversal symmetry. On the basis of topological classification, we also discuss possible other skin effects in arbitrary dimensions. Our work provides a unified understanding about the bulk-boundary correspondence and the skin effects in non-Hermitian systems.
\end{abstract}

\maketitle

Recently, non-Hermitian Hamiltonians~\cite{Hatano-96, Hatano-97, Bender-98, Bender-02, Bender-review, Konotop-review, Christodoulides-review} have been extensively studied in open classical~\cite{Makris-08, Klaiman-08, Guo-09, Ruter-10, Lin-11, Regensburger-12, Peng-14} and quantum~\cite{Lee-14, Nakagawa-18, Li-19, Wu-19, Yamamoto-19, Xiao-18} systems as well as disordered or correlated solids with finite-lifetime quasiparticles~\cite{Kozii-17, Zyuzin-18, Yoshida-18, Yoshida-19, Bergholtz-19, Kimura-19, McClarty-19}. In particular, much research has focused on distinctive characteristics of non-Hermitian topological phases~\cite{Rudner-09, Hu-11, Esaki-11, Schomerus-13, Poli-15, Zeuner-15, Zhen-15-exp, Malzard-15, Weimann-17-exp, Xu-17, Leykam-17, Xiao-17-exp, St-Jean-17-exp, Bahari-17-exp, Zhou-18-exp, Harari-18, Bandres-18-exp, Shen-18, Gong-18, Takata-18, KHGAU-19, Okugawa-19, Budich-19, Zhou-19, Longhi-19-AA, Cerjan-19-exp, Li-CHLee-19, KBS-19, KSUS-19, Lee-Vishwanath-19, Zeng-19, Herviou-entanglement-19, Chang-19}. The rich non-Hermitian topology is attributed to the complex-valued nature of the spectrum, which enables two types of complex-energy gaps~\cite{KSUS-19}: line gap and point gap. Since a non-Hermitian Hamiltonian with a line gap is continuously deformed to a Hermitian one without closing the line gap~\cite{KSUS-19}, topology for a line gap describes the persistence of conventional topological phases against non-Hermitian perturbations, which is relevant to topological lasers~\cite{St-Jean-17-exp, Bahari-17-exp, Zhou-18-exp, Harari-18, Bandres-18-exp}, for example. On the other hand, a non-Hermitian Hamiltonian with a point gap is allowed to be deformed to a unitary one~\cite{Gong-18, KSUS-19}. As a result, point-gapped topological phases cannot always be continuously deformed into any Hermitian counterparts; topology for a point gap is intrinsic to non-Hermitian systems in sharp contrast to a line gap. A point gap describes unique non-Hermitian topological phenomena such as localization transitions~\cite{Hatano-96, Hatano-97, Gong-18, Longhi-19-AA, Zeng-19} and emergence of exceptional points~\cite{Kozii-17, Zyuzin-18, Yoshida-18, Yoshida-19, Bergholtz-19, Kimura-19, Zhen-15-exp, Xu-17, Okugawa-19, Budich-19, Cerjan-19-exp, KBS-19}.

A hallmark of topological phases is the presence of the localized states at the boundaries as a result of nontrivial topology of the bulk~\cite{Kane-review, Zhang-review, Schnyder-Ryu-review}. Remarkably, non-Hermiticity alters the nature of the bulk-boundary correspondence (BBC)~\cite{Lee-16, MartinezAlvarez-18, Xiong-18, Kunst-18, YW-18, YSW-18, KSU-18, Clerk-18, Jin-19, Wang-19, Liu-19, Edvardsson-19, Lee-Thomale-19, Herviou-19, Kunst-19, Lee-Li-Gong-19, Yokomizo-19, Okuma-19, Longhi-19, Song-19, Imura-19, Rui-19, Schomerus-20, Zirnstein-19, Borgnia-19, Rui-19, Helbig-19-skin-exp, Ghatak-19-skin-exp, Xiao-19-skin-exp, Hofmann-19-skin-exp}. The critical distinction is the extreme sensitivity of the bulk to the boundary conditions, which is called the non-Hermitian skin effect~\cite{YW-18}. It accompanies the localization of bulk eigenstates as well as the dramatic difference of bulk spectra according to the boundary conditions, which forces us to redefine the bulk topology so as to be suitable for the open boundary condition~\cite{Kunst-18, YW-18, Kunst-19, Yokomizo-19}. The BBC persists in the presence of a line gap since non-Hermitian Hamiltonians with a line gap can be continuously deformed to Hermitian ones. However, the BBC for a point gap has still remained unclear. Since a point gap describes intrinsic non-Hermitian topology, the nature of the BBC may be disparate from the Hermitian counterpart. In fact, even when a point gap is open under the periodic boundary condition, it can be close under the open boundary condition~\cite{Gong-18, Xiong-18, Lee-Thomale-19}. Thus, the non-Hermitian skin effect obscures point-gap topology.

This Letter provides a unified understanding about the BBC and the skin effect in non-Hermitian systems. We show that the BBC holds even for a point gap in semi-infinite systems with only one boundary. In finite systems with open boundaries, by contrast, we demonstrate that the point-gap topology inevitably induces the non-Hermitian skin effect and results in the absence of topologically protected boundary states due to a point gap. On the basis of such a topological origin, new types of the skin effects are revealed, including the $\mathbb{Z}_{2}$ skin effect protected by time-reversal symmetry. We also elucidate the relationship between point and line gaps for the BBC.

%%%%%%%%%%
\paragraph{Bulk-boundary correspondence in semi-infinite systems.---}A non-Hermitian Hamiltonian $H$ is defined to have a point gap if and only if its complex spectrum does not cross a reference point $E \in \mathbb{C}$, i.e., $\det \left( H - E \right) \neq 0$~\cite{Gong-18, KSUS-19}. The simplest nontrivial example of the point-gapped topological phases appears in one-dimensional systems with no symmetry. Whereas $\det \left( H - E\right)$ is always real for Hermitian $H$, it can be complex for non-Hermitian $H$, by which the following winding number $W \left( E \right) \in \mathbb{Z}$ is defined: 
\begin{equation}
W \left( E \right)
:= \int_{0}^{2\pi} \frac{dk}{2\pi \ii} \frac{d}{dk} \log \det \left( H \left( k \right) - E \right),
	\label{eq: 1D class A - winding}
\end{equation}
where $H \left( k \right)$ is the non-Hermitian Bloch Hamiltonian in momentum space with the finite number of bands ($k \in \left[ 0, 2\pi \right]$). Topological phases are absent in one-dimensional Hermitian systems without symmetry protection~\cite{Kane-review, Zhang-review, Schnyder-Ryu-review}; the point-gap topology characterized by $W \left( E \right)$ is intrinsic to non-Hermitian systems.

Corresponding to $W \left( E \right) \neq 0$, the boundary modes with the eigenenergy $E$ can appear in semi-infinite systems with only one boundary. Suppose the non-Hermitian system has a boundary on the left but no boundary on the right (the same semi-infinite boundary condition is chosen below unless otherwise stated). An important observation is that the Hermitian Hamiltonian $\tilde{H}$ is obtained by~\cite{Gong-18, KSUS-19}
\begin{equation}
\tilde{H} := \left( \begin{array}{@{\,}cc@{\,}} 
	0 & H-E \\
	H^{\dag} - E^{*} & 0 \\ 
	\end{array} \right).
		\label{eq: extended - Hermitian}
\end{equation}
Under the periodic boundary condition, when a point gap is open for non-Hermitian $H \left( k \right)$, a real energy gap is also open for Hermitian $\tilde{H} \left( k \right)$, and vice versa. In addition, $\tilde{H}$ respects additional chiral symmetry by construction: $\Gamma \tilde{H} \Gamma^{-1} = - \tilde{H}$ with $\Gamma := \sigma_{z}$. As a result of the conventional BBC for Hermitian Hamiltonians, $\tilde{H}$ with the semi-infinite boundary possesses topologically protected zero modes localized at the boundary~\cite{Kane-review, Zhang-review, Schnyder-Ryu-review} in a similar manner to the Su-Schrieffer-Heeger model~\cite{SSH-79}. The corresponding topological invariant coincides with $W \left( E \right)$ in Eq.~(\ref{eq: 1D class A - winding}). For $W \left( E \right) < 0$, there appear boundary modes $\left( 0~\ket{E} \right)^{T}$ with negative chirality [i.e., $\Gamma \left( 0~\ket{E} \right)^{T} = - \left( 0~\ket{E} \right)^{T}$], which implies that $\ket{E}$ is a right eigenstate of non-Hermitian $H$ (i.e., $H \ket{E} = E \ket{E}$) localized at the boundary. For $W \left( E \right) > 0$, on the other hand, the boundary modes $\left( \ket{E}~0 \right)^{T}$ have positive chirality [i.e., $\Gamma \left( \ket{E}~0 \right)^{T} = + \left( \ket{E}~0 \right)^{T}$], which in turn implies that $\ket{E}$ is a right eigenstate of $H^{\dag}$, i.e., a left eigenstate of $H$ (i.e., $\bra{E} H = \bra{E}E$)~\cite{Brody-14}. 

The above discussion is valid for arbitrary $E \in \mathbb{C}$ in the complex-energy plane satisfying $W \left( E \right) \neq 0$. Thus, in semi-infinite systems $H_{\rm SIBC}$, the infinite number of boundary modes with eigenenergies $E$ emerges as a result of the nontrivial winding number $W \left( E \right) \neq 0$. This conclusion leads to the following theorem (index theorem in spectral theory~\cite{Blackadar, Higson-Roe, Bottcher, Trefethen, index, supplement}):\\

\begin{figure}[t]
\begin{center}
　　　\includegraphics[width=8cm,angle=0,clip]{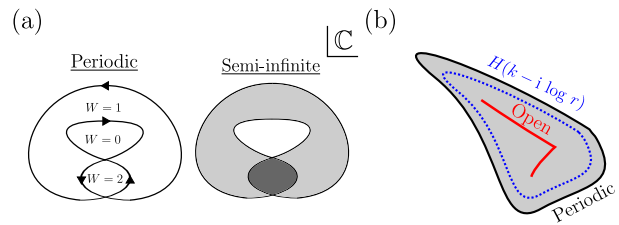}
　　　\caption{Complex spectra of non-Hermitian systems with periodic, open, and semi-infinite boundaries. (a)~A semi-infinite system possesses the infinite number of boundary modes due to the nonzero winding number $W \neq 0$ in the corresponding periodic system. (b)~The spectrum of a semi-infinite system shrinks through the imaginary gauge transformation, resulting in an arc of the open-boundary system.}
　　　\label{fig: Toeplitz}
\end{center}
\end{figure}

%\noindent
{\bf Theorem I}~~
Let $\sigma \left( H(k) \right)$ be the spectrum of $H\left(k \right)$ with $k\in [0,2\pi]$, which forms closed curves in the complex-energy plane (Fig.~\ref{fig: Toeplitz}). Then, the spectrum of semi-infinite $H_{\rm SIBC}$ with only one boundary is equal to $\sigma \left( H(k) \right)$ together with the whole area of $E \in \mathbb{C}$ enclosed by $\sigma \left( H(k) \right)$ with $W \left( E\right) \neq 0$. For $W \left( E \right) < 0$ ($W \left( E \right) > 0$), $\ket{E}$ is a right (left) eigenstate of $H_{\rm SIBC}$ localized at the boundary [i.e., $H_{\rm SIBC} \ket{E} = E \ket{E}$ ($\bra{E} H_{\rm SIBC} = \bra{E} E$)].\\

Theorem~I is illustrated with the Hatano-Nelson model~\cite{Hatano-96, Hatano-97} without disorder, which is given by 
\begin{equation}
H^{\rm (HN)} := \sum_i \left[ \left( t+g \right) c^{\dagger}_{i+1} c_i + \left( t-g \right)c^{\dagger}_{i} c_{i+1} \right]   
\label{eq: Hatano-Nelson}
\end{equation}
with $t > 0$ and $g \in \mathbb{R}$. The spectrum of the Bloch Hamiltonian $H^{\rm (HN)} \left( k \right) = \left( t+g \right) e^{\ii k} + \left( t-g \right) e^{-\ii k}$ forms an ellipse in the complex-energy plane, and we have $W \left( E \right) = \mathrm{sgn} \left( g\right)$ for $E \in \mathbb{C}$ inside this ellipse. In fact, the hopping from right to left dominates that from left to right for $g < 0$, which leads to the emergence of the boundary modes~\cite{supplement}.

%%%%%%%%%%
\paragraph{Skin effect as point-gap topology.\,---}The above discussion breaks down in finite systems with open boundaries. In fact, the infinite number of boundary modes is impossible in finite systems. Furthermore, an additional boundary condition is imposed because of the other boundary, which may forbid some of the boundary states appearing in semi-infinite systems. For example, the spectrum of the Hatano-Nelson model $H^{\rm (HN)}_{\rm OBC}$ with open boundaries forms not a loop but a line on the real axis in the complex-energy plane, which signals the non-Hermitian skin effect. In fact, using an imaginary gauge transformation~\cite{Hatano-96, Hatano-97, YW-18, Lee-Thomale-19}
\begin{equation}
V_r^{-1} c_i^{\dagger} V_r = r^i c_i^{\dagger},\quad
V_r^{-1} c_i V_r = r^{-i} c_i,\quad
\left( 0 < r < \infty \right)
	\label{eq: imaginary_gauge}
\end{equation}
we have a Hermitian Hamiltonian $\bar{H} := V_{r}^{-1} H^{\rm (HN)}_{\rm OBC} V_{r}$ for $r := \sqrt{\left| \left( t-g \right)/\left(t+g \right) \right|}$. Here, Eq.~(\ref{eq: imaginary_gauge}) shifts the momentum from $k$ to $k-\ii \log r$. Since this similarity transformation does not change the spectrum, $H^{\rm (HN)}_{\rm OBC}$ has the entirely real spectrum and hence no longer retains the point gap. Saliently, such a non-Hermitian skin effect is a general non-Hermitian topological phenomenon as a direct consequence of point-gap topology, as summarized in the following theorem:\\

{\bf Theorem II}~~Finite $H_{\rm OBC}$ with open boundaries is always topologically trivial in terms of a point gap. Consequently, if $H \left( k \right)$ under the periodic boundary condition is point-gapped topological, the non-Hermitian skin effect inevitably occurs with a topological phase transition.\\

To see this theorem, we begin with
\begin{equation}
\lim_{N \to \infty} \sigma \left( H_{\rm OBC} \right) \subset \sigma \left( H_{\rm SIBC} \right),
	\label{eq: inclusion OBC-SIBC}
\end{equation}
where $\sigma \left( H_{\rm OBC} \right)$ is the spectrum of a non-Hermitian system $H_{\rm OBC}$ with open boundaries and $N$ unit cells, and $\sigma \left( H_{\rm SIBC} \right)$ is the spectrum of the corresponding semi-infinite system $H_{\rm SIBC}$. In fact, an approximate eigenstate of $H_{\rm SIBC}$ can be obtained from an eigenstate of $H_{\rm OBC}$, which becomes an exact eigenstate for $N \to \infty$~\cite{supplement}. The contrary is not always true: even if an approximate eigenstate of $H_{\rm OBC}$ is constructed from an eigenstate of $H_{\rm SIBC}$, it is not necessarily an exact eigenstate of $H_{\rm OBC}$.

A crucial step is again the imaginary gauge transformation: $H_{\rm OBC} \rightarrow V_{r}^{-1} H_{\rm OBC} V_{r}$ and $H_{\rm SIBC} \rightarrow V_r^{-1} H_{\rm SIBC} V_r$ with $r \in \left( 0, \infty \right)$. For each transformation, we still have the inclusion in Eq.~(\ref{eq: inclusion OBC-SIBC}):
\begin{equation}
\lim_{N \to \infty} \sigma \left( V_{r}^{-1} H_{\rm OBC} V_{r} \right) \subset \sigma \left( V^{-1}_{r} H_{\rm SIBC} V_{r} \right).
	\label{eq: inclusion OBC-SIBC-2}
\end{equation}
This imaginary gauge transformation does not change the spectrum of $H_{\rm OBC}$. However, it changes the spectrum of $H_{\rm SIBC}$ since $H_{\rm SIBC}$ has no boundary on the right because of the semi-infinite nature [Fig.~\ref{fig: Toeplitz}\,(b)]. In fact, $H \left( k \right)$ changes to $H \left( k-\ii\log r \right)$ through $V_{r}$. Nevertheless, Eq.~(\ref{eq: inclusion OBC-SIBC-2}) implies that the transformed semi-infinite spectrum includes the spectrum of $H_{\rm OBC}$ for any transformation $V_{r}$. Thus, we have
\begin{equation}
\lim_{N \to \infty} \sigma \left( H_{\rm OBC} \right) \subset \bigcap_{r \in \left( 0, \infty \right)} \sigma \left( V^{-1}_{r} H_{\rm SIBC} V_{r} \right).
	\label{eq: inclusion OBC-SIBC-3}
\end{equation}

Because of Theorem~I, when $H \left( k \right)$ has a point gap and $W \left( E \right) < 0$ ($W \left( E \right) > 0$), right (left) boundary modes with eigenenergy $E$ appear in the semi-infinite system. Let us choose an appropriate imaginary gauge $V_{r}$ such that these boundary modes are transformed to delocalized bulk modes. Then, $E$ is on the edges of $\sigma \left( V_{r}^{-1} H_{\rm SIBC} V_{r}\right)$, whereas it is originally located inside $\sigma \left( H_{\rm SIBC} \right)$. Thus, the intersection of $\sigma \left( H_{\rm SIBC} \right)$ and $\sigma \left( V_{r}^{-1} H_{\rm SIBC} V_{r}\right)$ is strictly smaller than $\sigma \left( H_{\rm SIBC} \right)$~\cite{supplement}. Repeating this procedure for all $V_{r}$ with $r \in \left( 0, \infty \right)$, the right-hand side of Eq.~(\ref{eq: inclusion OBC-SIBC-3}) reaches an open curve or a topologically trivial area of which interior satisfies $W(E)=0$, otherwise a contradiction arises~\cite{supplement}. Since this region includes $\lim_{N \to \infty} \sigma \left( H_{\rm OBC} \right)$ because of Eq.~(\ref{eq: inclusion OBC-SIBC-3}), $H_{\rm OBC}$ is also topologically trivial and different from $H \left( k \right)$ with nontrivial topology. Furthermore, $\sigma \left( H_{\rm OBC} \right)$ is indeed distinct from $\sigma \left( H \left( k \right) \right)$, which implies the inevitable occurrence of the non-Hermitian skin effect due to the point-gap topology.

Remarkably, Refs.~\cite{Kunst-18, YW-18, Kunst-19, Yokomizo-19} determine the conditions for the spectra of open-boundary systems and develop the non-Bloch band theory of non-Hermitian systems. Their conditions are actually equivalent to the set in the right-hand side of Eq.~(\ref{eq: inclusion OBC-SIBC-3})~\cite{supplement}. An observation similar to our Theorem~II is also made in Ref.~\cite{Lee-Thomale-19}, which is made rigorous by our results. Moreover, we identify the non-Hermitian skin effect as the point-gap topology~\cite{converse}. Such a topological origin constitutes a universal feature of the non-Hermitian skin effect. Furthermore, new types of the skin effects---symmetry-protected skin effects---are discovered, as illustrated below.

%%%%%%%%%%
\paragraph{$\mathbb{Z}_2$ non-Hermitian skin effect.\,---}The point-gap topology and the corresponding skin effect are enriched by symmetry. Here, we consider time-reversal symmetry defined in terms of transposition~\cite{KSUS-19}:
\begin{align}
T H^{T}\left( k \right) T^{-1} = H \left( -k \right),\quad TT^{*} = -1,
	\label{eq: TRS}
\end{align}
where $T$ is a unitary operator. This symmetry is fundamental as reciprocity in non-Hermitian spinful systems and naturally appears, for example, in mesoscopic systems~\cite{Beenakker-97, Beenakker-15} and open quantum systems~\cite{Hamazaki-19, Lieu-19, Breuer}.

In conventional quantum spin Hall insulators, the integer Chern number vanishes but the Kane-Mele $\mathbb{Z}_2$ one becomes nontrivial because of time-reversal symmetry~\cite{Kane-review, Zhang-review, Schnyder-Ryu-review}. Similarly, Eq.~(\ref{eq: TRS}) trivializes the winding number in Eq.~(\ref{eq: 1D class A - winding}), but instead, it supplies a $\mathbb{Z}_2$ invariant. The $\mathbb{Z}_{2}$ topological invariant $\nu \left( E \right) \in \{ 0, 1\}$ for a reference point $E \in \mathbb{C}$ is given by~\cite{KSUS-19}
\begin{align}
&\left( -1 \right)^{\nu \left( E \right)} := \mathrm{sgn} \left\{
\frac{ \mathrm{Pf} \left[ \left( H \left( \pi \right) - E \right) T \right] }{ \mathrm{Pf} \left[ \left( H \left( 0 \right) - E \right) T \right] } \right. \nonumber \\
&\qquad\left. \times \exp \left[ 
-\frac{1}{2} \int_{k=0}^{k=\pi} d \log \det \left[ \left( H \left( k \right) - E \right) T \right]
\right] \right\}.
	\label{eq: AII-dag Z2 inv}
\end{align}

Corresponding to the $\mathbb{Z}_{2}$ topological invariant $\nu \left( E \right)$, we have an index theorem similar to Theorem~I for semi-infinite systems~\cite{supplement}. A clear distinction from Theorem~I is the Kramers degeneracy due to Eq.~(\ref{eq: TRS})~\cite{Esaki-11, Zhou-19, KSUS-19}. The extended Hermitian Hamiltonian $\tilde{H}$ in Eq.~(\ref{eq: extended - Hermitian}) respects time-reversal symmetry as well as the additional chiral symmetry $\Gamma$, analogous to time-reversal-invariant topological superconductors ~\cite{Qi-09, QHZ-10, SR-11, Budich-13}. 
The index theorem states that the semi-infinite system $\tilde{H}$ hosts an odd number of boundary Majorana Kramers pairs for each $E$ with $\nu \left( E \right) = 1$. In terms of the original non-Hermitian Hamiltonian $H$, the Kramers pair reduces to a pair of right and left eigenstates of $H$ localized at the same boundary. Using the transposition version of time reversal in Eq. (\ref{eq: TRS}), we can convert the left eigenmode into a right one in the oppositely extended semi-infinite system (i.e., semi-infinite system with a boundary only on the right). As a result, finite systems with open boundaries host localized modes at both ends, as explicitly shown in the following model.

\begin{figure}[t]
\begin{center}
　　　\includegraphics[width=8cm,angle=0,clip]{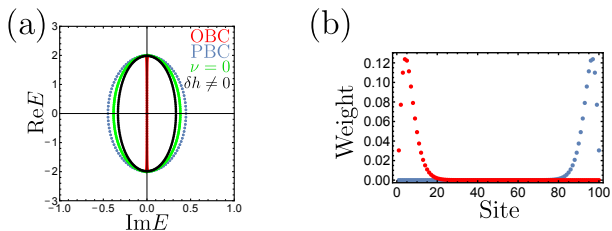}
　　　\caption{$\mathbb{Z}_{2}$ non-Hermitian skin effect. (a)~Energy spectra of the non-Hermitian Hamiltonian in Eq.~(\ref{eq: 1D AII-dag}) under the various boundary conditions ($t=1$, $g=0.3$, $\Delta=0.2$, $\bm{\delta}=(1,1,1)\times10^{-2}$, $\delta h=10^{-3}$, $N=100$). (b)~Kramers doublet with $E=1.948$, one of which is localized at the left boundary and the other at the right boundary.}
　　　\label{fig: 1D AII-dag}
\end{center}
\end{figure}

We recall that a quantum spin Hall insulator~\cite{Kane-Mele-05-Z2, Kane-Mele-05-QSH} can be constructed from a pair of time-reversed quantum Hall insulators~\cite{Haldane-88} with the spin-orbit coupling. Similarly, combining the Hatano-Nelson model $H^{\rm (HN)} \left( k\right)$ in Eq.~(\ref{eq: Hatano-Nelson}) and its time-reversed partner $( H^{\rm (HN)} )^{T} \left( -k \right)$, we have a canonical model that exhibits the $\mathbb{Z}_{2}$ skin effect:
\begin{align}
H \left( k \right)
&= \left( \begin{array}{@{\,}cc@{\,}} 
	H^{\rm (HN)} \left( k \right) & 2\Delta \sin k \\
	2\Delta \sin k & ( H^{\rm (HN)} )^{T} \left( -k \right) \\ 
	\end{array} \right) \nonumber \\
&= 2 t \cos k + 2 \Delta \left( \sin k \right) \sigma_{x} + 2\ii g \left( \sin k \right) \sigma_{z},
	\label{eq: 1D AII-dag}
\end{align}
with $t, g, \Delta \geq 0$. It indeed respects time-reversal symmetry with $T = \ii \sigma_{y}$, and its spectrum is given as $E_{\pm} \left( k \right) = 2t \cos k \pm 2\ii \sqrt{g^2-\Delta^2} \sin k$. Thus, $H \left( k \right)$ for $g>\Delta$ retains a point gap. Since it can be continuously deformed to $H \left( k \right)$ with $t=g, \Delta=0$ while keeping the point gap, the $\mathbb{Z}_2$ invariant in Eq.~(\ref{eq: AII-dag Z2 inv}) is obtained as $\nu \left( E \right) = 1$ when $E$ is in the area enclosed by $\sigma \left( H(k) \right)$.

The spectrum of Eq.~(\ref{eq: 1D AII-dag}) is shown in Fig.~\ref{fig: 1D AII-dag}\,(a). The open-boundary spectrum is clearly different from the periodic-boundary counterpart, which indicates the non-Hermitian skin effect. Each complex eigenenergy consists of a Kramers pair, one of which is localized at the left boundary and the other at the right boundary~[Fig.~\ref{fig: 1D AII-dag}\,(b)]. Because of the $\mathbb{Z}_{2}$ nature, the point-gap topology becomes trivial and no skin effect occurs if the two nontrivial systems are stacked. Figure~\ref{fig: 1D AII-dag}\,(a) also shows the spectrum of such a stacked system
\begin{align}
H^{\mathrm{stack}} \left( k \right)= \left( \begin{array}{@{\,}cc@{\,}} 
	H \left( k \right) & \ii \bm{\delta} \cdot \bm{\sigma} \\
	-\ii \bm{\delta} \cdot \bm{\sigma} & H \left( k \right) \\ 
	\end{array} \right),
\end{align}
where the off-diagonal terms are symmetry-preserving couplings. Consistently, the non-Hermitian skin effect no longer survives.

Since the $\mathbb{Z}_2$ skin effect is topologically protected by time-reversal symmetry, it breaks down by a symmetry-breaking perturbation including $\left(\delta h \right)\sigma_z$ [Fig.~\ref{fig: 1D AII-dag}\,(a)]. In particular, such a local perturbation, which does not connect the ends, may be infinitesimal for the breakdown of the skin effect~\cite{Okuma-19}. This local infinitesimal instability is unique to symmetry-protected non-Hermitian skin effects.%~\cite{infinitesimal}.

%%%%%%%%%%
\paragraph{Bulk-boundary correspondence in finite systems.\,---}

General theories on the BBC in non-Hermitian systems have recently been developed~\cite{Kunst-18, YW-18, Kunst-19, Yokomizo-19}. These theories implicitly consider non-Hermitian topology for a line gap~\cite{KSUS-19}. A non-Hermitian Hamiltonian $H$ is defined to have a line gap if and only if its spectrum does not cross a reference line in the complex-energy plane. The modified BBC persists because a non-Hermitian Hamiltonian with a line gap can be continuously deformed to a Hermitian one~\cite{KSUS-19}. On the other hand, we develop a theory of the BBC for a point gap, which complements Refs.~\cite{Kunst-18, YW-18, Kunst-19, Yokomizo-19}.

A prototypical example is a non-Hermitian extension of the Su-Schrieffer-Heeger model~\cite{SSH-79} with asymmetric hopping~\cite{Lee-16, Kunst-18, YW-18, Yokomizo-19, supplement}. It exhibits the skin effect under the open boundary condition due to the point-gap topology characterized by Eq.~(\ref{eq: 1D class A - winding}) under the periodic boundary condition~\cite{supplement}. %~\cite{SSH-point}. 
Still, a line gap can be open and the corresponding topological invariant protected by sublattice symmetry can be well defined under the open boundary condition. As a result, topologically protected zero modes can emerge because of this line-gap topology.

Importantly, point and line gaps are not necessarily independent of each other. In fact, if a line gap is open, a point gap is also open with a reference point on the reference line. Hence, a reminiscence of line-gap topology may survive in the presence of a point gap even if the line gap is closed. A prime example includes non-Hermitian superconductors in one dimension without time-reversal symmetry. In this case, particle-hole symmetry
\begin{equation}
C H^{T} \left( k \right) C^{-1} =- H \left( -k \right),\quad CC^{*} = +1
\end{equation}
makes zero energy a special point in the complex-energy plane in contrast to time-reversal symmetry. As a result, non-Hermitian systems have the $\mathbb{Z}_{2}$ topological phases for both point and line gaps, and their topological invariants coincide with each other~\cite{KSUS-19}. The Majorana zero modes in Hermitian topological superconductors survive as long as the point gap at $E = 0$ is open. Correspondingly, an index theorem states the emergence of the zero modes localized at the boundary~\cite{supplement}. A concrete model of such a non-Hermitian $s$-wave topological superconductor is provided in Ref.~\cite{Okuma-19}. To characterize this type of point-gap topology in a general manner, Refs.~\cite{Shiozaki-19, supplement} classify the homomorphisms from line-gap topology to point-gap topology for all the 38-fold internal symmetry class in arbitrary spatial dimensions.

%%%%%%%%%%
\paragraph{Higher-dimensional skin effects.\,---}

By contrast, point-gap topology can be nontrivial even if line-gap topology is trivial. For example, whereas line-gap topology is absent in one dimension with and without time-reversal symmetry~\cite{KSUS-19}, the point-gap topology characterized by Eqs.~(\ref{eq: 1D class A - winding}) and (\ref{eq: AII-dag Z2 inv}) is present. As shown in this Letter, such intrinsic point-gap topology in finite systems leads to not the BBC but the skin effect. References~\cite{Shiozaki-19, supplement} also classify the non-Hermitian topology unique to a point gap. This classification allows us to know possible types of symmetry-protected skin effects for general symmetry classes and arbitrary dimensions. Like surface Dirac fermions in topological insulators, higher-dimensional skin modes appear in any boundary of the system under a proper boundary condition~\cite{weak}.

For example, a two-dimensional variant of the $\mathbb{Z}_2$ skin effect is investigated in Ref.~\cite{supplement}. There, skin modes coexist with bulk modes under the open boundary condition in one direction and the periodic boundary condition in the other direction, which is the ``proper boundary condition" in this system. Remarkably, only $\mathcal{O}\,( L )$ skin modes appear from all the $\mathcal{O}\,( L^{2} )$ modes in this model ($L$ denotes the length in one direction), which is unfeasible for the skin effects in one dimension.

%%%%%%%%%%
\paragraph{Discussion.\,---}

The non-Hermitian skin effect has recently been observed in electrical circuits~\cite{Helbig-19-skin-exp, Hofmann-19-skin-exp}, a mechanical metamaterial~\cite{Ghatak-19-skin-exp}, and quantum walk~\cite{Xiao-19-skin-exp}, all of which we identify are intrinsic non-Hermitian topological phenomena. Beyond the observed one, this Letter predicts novel types of skin effects enabled by symmetry protection. It merits further research to investigate a variety of symmetry-protected non-Hermitian skin effects and their new physics.

%%%%% Acknowledgement %%%%%
We thank Yosuke Kubota for helpful discussions. This work was supported by a Grant-in-Aid for Scientific Research on Innovative Areas ``Topological Materials Science" (KAKENHI Grant No.~JP15H05855) from the Japan Society for the Promotion of Science (JSPS). This work was also supported by JST CREST Grant No.~JPMJCR19T2, Japan. N.O. was supported by KAKENHI Grant No.~JP18J01610 from the JSPS. K.K. was supported by KAKENHI Grant No.~JP19J21927 from the JSPS. K.S. was supported by PRESTO, JST (Grant No.~JPMJPR18L4). M.S. was supported by KAKENHI Grant No.~JP17H02922 from the JSPS.

%%%%% Note %%%%%
%\smallskip
{\it Note added.\,---\,}After completion of this work, we became aware of a recent related work~\cite{Zhang-19}.

%\bibliographystyle{apsrev4-1}
%\bibliography{NH-topo-BBC}
%merlin.mbs apsrev4-1.bst 2010-07-25 4.21a (PWD, AO, DPC) hacked
%Control: key (0)
%Control: author (72) initials jnrlst
%Control: editor formatted (1) identically to author
%Control: production of article title (-1) disabled
%Control: page (0) single
%Control: year (1) truncated
%Control: production of eprint (0) enabled
%

%%%%% Supplemental Material %%%%%
\widetext
\pagebreak

\renewcommand{\theequation}{S\arabic{equation}}
\renewcommand{\thefigure}{S\arabic{figure}}
\renewcommand{\thetable}{S\arabic{table}}
\setcounter{equation}{0}
\setcounter{figure}{0}
\setcounter{table}{0}

\begin{center}
{\bf \large Supplemental Material for %\\ \smallskip 
``Topological Origin of Non-Hermitian Skin Effects"}
\end{center}

%%%%%%%%%%
\section{SI.~Toeplitz index theorem}

The bulk-boundary correspondence (BBC) and the skin effect in non-Hermitian one-dimensional systems are understood on the basis of spectral theory of Toeplitz matrices and operators~\cite{Trefethen, Bottcher}. A Toeplitz matrix $H_{\mathrm{OBC}}$ and a Toeplitz operator $H_{\mathrm{SIBC}}$ are respectively an $N \times N$ matrix and an operator in semi-infinite space in which each descending diagonal from left to right is constant:
\begin{align}
H_{\mathrm{OBC}} :=
\begin{pmatrix}
	a_0 & a_{-1} & \cdots & a_{1-N}\\
	a_1 & a_0 & \ddots & \vdots\\
	\vdots&\ddots&\ddots& a_{-1} \\
	a_{N-1}& \cdots & a_1 & a_0
\end{pmatrix},\quad
H_{\mathrm{SIBC}} :=
\begin{pmatrix}
	a_0&a_{-1}&\cdots&\cdots\\
	a_1&a_0&\ddots&\\
	\vdots&\ddots&\ddots&\ddots\\
	\vdots&&\ddots&\ddots
\end{pmatrix},\quad
a_{i} \in \mathbb{C}.
\end{align}
The simplest example is the clean Hatano-Nelson model~\cite{Hatano-96, Hatano-97}, which is defined by $a_{\pm1} := t \pm g$ ($t,g\in\mathbb{R}$) and $a_{i} := 0$ for $i \neq \pm 1$. We define the symbol of the Toeplitz matrices and operators as
\begin{align}
H \left( \beta \right) := \sum_{i} a_{i} \beta^{i},\quad \beta \in \mathbb{C}.
	\label{symbol}
\end{align}
If a Toeplitz matrix is periodic, i.e., $a_{i} = a_{i-N}$, it is called a circulant matrix. The spectrum $\sigma\,(H_{\rm PBC})$ of a circulant matrix $H_{\rm PBC}$ is given in terms of the symbol as
\begin{align}
\sigma\,(H_{\rm PBC}) = H \left( \mathbb{T}_N \right),
\quad \mathbb{T}_N := \{\beta\,|\,\beta^N=1\}.
\end{align}
A circulant matrix describes the system with periodic boundaries, and its spectrum is nothing but the Fourier transform in momentum space.

Similarly, a Laurent operator is defined as an infinite operator that has the same form as the corresponding Toeplitz matrix and operator. The spectrum $\sigma\,(H_{\mathrm{Laurent}})$ of a Laurent operator $H_{\mathrm{Laurent}}$ is given as
\begin{align}
\sigma\,(H_{\mathrm{Laurent}})=H \left( \mathbb{T} \right),\quad
\quad \mathbb{T} := \{\beta\,|\,|\beta|=1\}.
\end{align}
Thus, the spectrum of a Laurent operator includes that of the corresponding circulant matrix, and we have $\lim_{N \to \infty} \sigma\,(H_{\rm PBC}) = \sigma\,(H_{\mathrm{Laurent}})$.

In contrast to circulant matrices and Laurent operators, Toeplitz matrices $H_{\mathrm{OBC}}$ and operators $H_{\mathrm{SIBC}}$ lack periodicity and describe finite systems with open boundaries and semi-infinite systems with one boundary, respectively. Hence, the spectrum of a Toeplitz matrix or operator can be different from $\sigma\,(H_{\rm PBC})$ and $\sigma\,(H_{\mathrm{Laurent}})$ in the presence of non-Hermiticity (nonnormality), which is the non-Hermitian skin effect~\cite{YW-18}. There is a useful theorem to characterize the spectrum of a Toeplitz operator~\cite{Trefethen}:

\begin{itemize}
\item[] \textbf{Theorem~S1}~~Let $H_{\mathrm{SIBC}}$ be a Toeplitz operator with a continuous symbol $H \left( \beta \right)$. Then, the spectrum $\sigma \left( H_{\mathrm{SIBC}} \right)$ is equal to $H \left( \mathbb{T} \right)$ together with all the points $E \in \mathbb{C}$ enclosed by the curve $H \left( \mathbb{T} \right)$ with the nonzero winding number $W \left(E \right) \neq 0$.
\end{itemize}

Here, the winding number $W \left( E \right)$ of $H \left( \beta \right)$ for a reference point $E \in \mathbb{C}$ is defined as
\begin{equation}
W \left( E \right)
:= \oint_{C} \frac{d\beta}{2\pi \ii} \frac{d}{d\beta} \log \left( H \left( \beta \right) - E \right),
\end{equation} 
where the integral is taken on a loop $C$ in the complex plane that encloses a reference point $E \in \mathbb{C}$. This winding number is identical to the topological invariant of a generic non-Hermitian Hamiltonian with a point gap~\cite{Gong-18, KSUS-19}. Theorem S1 is equivalent to the Toeplitz index theorem~\cite{Bottcher}:
\begin{align}
	\dim \mathrm{Ker}\,H_{\mathrm{SIBC}} - \dim \mathrm{Coker}\,H_{\mathrm{SIBC}} = - W \left( 0 \right),
\end{align}
where we denote $\mathrm{Ker}\,H_{\mathrm{SIBC}} := \{x\in X\ |\ H_{\mathrm{SIBC}}\ x=0 \}$, $\mathrm{Coker}\,H_{\mathrm{SIBC}} := X/\mathrm{Im}\,H_{\mathrm{SIBC}}$, and $\mathrm{Im}\,H_{\mathrm{SIBC}}:=\{H_{\mathrm{SIBC}}\ x\ |\ x\in X\}$ with vector space $X$ on which $H_{\mathrm{SIBC}}$ acts. The corresponding eigenvectors are determined by the following theorem~\cite{Trefethen}:

\begin{itemize}
\item[] \textbf{Theorem~S2}~~For $W \left( E \right) < 0$, $E$ is an eigenvalue of $H_{\mathrm{SIBC}}$ and the amplitude of the corresponding (right) eigenvector decreases for $i \rightarrow \infty$. If $H \left( \beta \right)$ is a rational function, it decreases exponentially.
\end{itemize}

This theorem characterizes eigenstates of non-Hermitian Hamiltonians. If we assume the locality of the Hamiltonian, i.e., the sum of the symbol $H \left( \beta \right)$ in Eq.~(\ref{symbol}) is finite, $H \left( \beta \right)$ is clearly a rational function. In this situation, exponentially-localized boundary modes can also be eigenstates of the semi-infinite Hamiltonian, in addition to the plane waves characterized by $H \left( \mathbb{T} \right)$. In fact, for $W \left( E \right) < 0$, such boundary modes are indeed eigenstates. For $W \left( E \right) > 0$, on the other hand, $E$ is no longer an eigenenergy since there is no boundary on the right because of the semi-infinite nature. Instead, $E$ is an eigenenergy of $H_{\mathrm{SIBC}}^T$. The corresponding eigenstate is a boundary mode of the semi-infinite Hamiltonian with the right boundary. Remarkably, the spectrum $\sigma \left( H_{\mathrm{SIBC}} \right)$ of a Toeplitz operator $H_{\mathrm{SIBC}}$ is the set of $E$ in which $E - H_{\mathrm{SIBC}}$ is not invertible. As a consequence, $\sigma \left( H_{\mathrm{SIBC}} \right)$ in infinite-dimensional vector space is not necessarily identical to $\sigma \left( H_{\mathrm{OBC}} \right)$ or $\sigma~( H_{\rm PBC} )$ in finite-dimensional vector space.

The above theorems can be generalized to essentially normal operators $H$~\cite{Blackadar, Higson-Roe}. Here, an essentially normal operator is defined as an operator in which $[H, H^\dagger]$ is compact. Physically, non-Hermitian local Hamiltonians are essentially normal. Consequently, if the system respects translation invariance except for the boundaries, the above theorems are applicable.

%%%%%%%%%%
\section{SII.~Index theorems in the presence of symmetry}

\subsection{Class $\text{AII}^{\dag}$ (time-reversal symmetry)}

Non-Hermitian Hamiltonians $H$ in class $\text{AII}^{\dag}$ respect time-reversal symmetry~\cite{KSUS-19}:
\begin{align}
T H^{T} \left( k \right) T^{-1} = H \left( -k \right),\quad TT^{*} = -1,
\end{align}
where $T$ is a unitary operator. In the presence of a point gap for a reference point $E \in \mathbb{C}$, the following $\mathbb{Z}_{2}$ topological invariant $\nu \left( E \right) \in \{ 0, 1\}$ is defined:
\begin{equation}
\left( -1 \right)^{\nu \left( E \right)} := \mathrm{sgn} \left\{
\frac{ \mathrm{Pf} \left[ \left( H \left( \pi \right) - E \right) T \right] }{ \mathrm{Pf} \left[ \left( H \left( 0 \right) - E \right) T \right] }
\times \exp \left[ 
-\frac{1}{2} \int_{k=0}^{k=\pi} d \log \det \left[ \left( H \left( k \right) - E \right) T \right]
\right] 
\right\}.
\end{equation}
In semi-infinite systems $H_{\rm SIBC}$, the index theorem reads
\begin{equation}
\# \left[ \mathrm{zero~modes~of}~\left( H_{\rm SIBC} - E \right)\right]
\equiv \nu \left( E \right) \quad \left( \mathrm{mod}~2 \right).
\end{equation}

To see this theorem, we consider the extended Hermitian Hamiltonian
\begin{equation}
\tilde{H} := \left( \begin{array}{@{\,}cc@{\,}} 
	0 & H-E \\
	H^{\dag} - E^{*} & 0 \\ 
	\end{array} \right),
	\label{seq: extended Hermitian}
\end{equation}
which belongs to symmetry class DIII~\cite{QHZ-10, SR-11, Budich-13} with chiral and time-reversal symmetries described by
\begin{equation}
\Gamma := \left( \begin{array}{@{\,}cc@{\,}} 
	1 & 0 \\
	0 & -1 \\ 
	\end{array} \right),\quad
\tilde{T} := \left( \begin{array}{@{\,}cc@{\,}} 
	0 & T \\
	T & 0 \\ 
	\end{array} \right) K,
		\label{seq: CS - TRS}
\end{equation}
where $K$ denotes complex conjugation. The conventional BBC for the Hermitian Hamiltonian $\tilde{H}$ states that $\tilde{H}$ has a Kramers pair of zero modes localized at the boundary for $\nu \left( E\right) = 1$. One of the zero modes has positive chirality (i.e., $\Gamma \ket{\varphi_{+}} = + \ket{\varphi_{+}}$) and the other has negative chirality (i.e., $\Gamma \ket{\varphi_{-}} = - \ket{\varphi_{-}}$), both of which are related by time reversal (i.e., $\tilde{T} \ket{\varphi_{-}} = \ket{\varphi_{+}}$). Thus, these zero modes are represented as
\begin{equation}
\ket{\varphi_{-}} = \left( \begin{array}{@{\,}c@{\,}} 
	0 \\
	\ket{E} \\
	\end{array} \right),\quad
\ket{\varphi_{+}}
=  \tilde{T} \ket{\varphi_{-}}
= \left( \begin{array}{@{\,}c@{\,}} 
	T \ket{E}^{*} \\
	0 \\
	\end{array} \right),
\end{equation}
which leads to
\begin{equation}
H \ket{E} = E \ket{E},\quad
H^{\dag} \left( T \ket{E}^{*} \right)
= E^{*} \left( T \ket{E}^{*} \right).
\end{equation}
These equations imply that $\ket{E}$ ($T \ket{E}^{*}$) is a right (left) eigenstate of the semi-infinite non-Hermitian Hamiltonian with $\nu \left( E\right) = 1$. Because of the $\mathbb{Z}_{2}$ nature of topology, only the parity of the boundary modes is relevant.

\subsection{Class D (particle-hole symmetry)}

Non-Hermitian Hamiltonians $H$ in class D respect particle-hole symmetry~\cite{KSUS-19}:
\begin{align}
C H^{T} \left( k \right) C^{-1} = - H \left( -k \right),\quad CC^{*} = +1,
\end{align}
where $C$ is a unitary operator. Particle-hole symmetry creates a pair of eigenenergies $\left( E, -E \right)$ and makes zero energy a special symmetric point in the complex-energy plane. Hence, we take a reference point as zero energy so that it will respect particle-hole symmetry. In the presence of the point gap, the following $\mathbb{Z}_{2}$ topological invariant $\nu \in \{ 0, 1\}$ is defined:
\begin{equation}
\left( -1 \right)^{\nu} := \mathrm{sgn} \left\{
\frac{ \mathrm{Pf} \left[ H \left( \pi \right) C \right] }{ \mathrm{Pf} \left[ H \left( 0 \right) C \right] }
\times \exp \left[ 
-\frac{1}{2} \int_{k=0}^{k=\pi} d \log \det \left[ H \left( k \right) C \right]
\right] 
\right\}.
\end{equation}
In semi-infinite systems $H_{\rm SIBC}$, the index theorem reads
\begin{equation}
\# \left[ \mathrm{zero~modes~of}~H_{\rm SIBC}\right]
\equiv \nu\quad \left( \mathrm{mod}~2 \right).
\end{equation}

To see this theorem, we consider the extended Hermitian Hamiltonian $\tilde{H}$ in Eq.~(\ref{seq: extended Hermitian}) with $E=0$, which belongs to symmetry class DIII~\cite{QHZ-10, SR-11, Budich-13} in a similar manner to non-Hermitian Hamiltonians in class $\text{AII}^{\dag}$. For $\tilde{H}$, particle-hole symmetry is described by
\begin{equation}
\tilde{C} := \left( \begin{array}{@{\,}cc@{\,}} 
	0 & C \\
	C & 0 \\ 
	\end{array} \right) K,
\end{equation}
and time-reversal symmetry is described as a combination of particle-hole and chiral symmetries by
\begin{equation}
\tilde{T} := \ii \Gamma \tilde{C}
= \ii \left( \begin{array}{@{\,}cc@{\,}} 
	0 & C \\
	-C & 0 \\ 
	\end{array} \right) K.
\end{equation}
Here, $\tilde{T}$ is chosen so that it will commute with $\tilde{C}$ (i.e., $\tilde{C}\tilde{T} = \tilde{T}\tilde{C}$). The BBC for the Hermitian Hamiltonian $\tilde{H}$ states that $\tilde{H}$ has a Kramers pair of zero modes localized at the boundary for $\nu = 1$. One of the zero modes has positive chirality (i.e., $\Gamma \ket{\varphi_{+}} = + \ket{\varphi_{+}}$) and the other has negative chirality (i.e., $\Gamma \ket{\varphi_{-}} = - \ket{\varphi_{-}}$), both of which are related by time reversal (i.e., $\tilde{T} \ket{\varphi_{-}} = \ket{\varphi_{+}}$). Thus, these zero modes are represented as
\begin{equation}
\ket{\varphi_{-}} = \left( \begin{array}{@{\,}c@{\,}} 
	0 \\
	\ket{0} \\
	\end{array} \right),\quad
\ket{\varphi_{+}}
=  \tilde{T} \ket{\varphi_{-}}
= \left( \begin{array}{@{\,}c@{\,}} 
	\ii C \ket{0}^{*} \\
	0 \\
	\end{array} \right),
\end{equation}
which leads to
\begin{equation}
H \ket{0} =
H^{\dag} \left( C \ket{0}^{*} \right)
= 0.
\end{equation}
These equations imply that $\ket{0}$ ($C \ket{0}^{*}$) is a right (left) eigenstate with zero energy of the semi-infinite non-Hermitian Hamiltonian with $\nu = 1$.

%%%%%%%%%%
\section{SIII.~Hatano-Nelson model}

We investigate the semi-infinite Hatano-Nelson model~\cite{Hatano-96, Hatano-97}
\begin{equation}
H^{\rm (HN)} \left( k \right)
= \left( t+g \right) e^{\ii k} + \left( t-g \right) e^{-\ii k}
\end{equation}
and determine its boundary modes. We assume $t > \left| g \right| \geq 0$ for the sake of simplicity. Suppose the system has a boundary on the left but no boundary on the right. If a state $\ket{E} = \left( \psi_{1}~\psi_{2}~\psi_{3}~\cdots\right)^{T}$ is an eigenstate with an eigenenergy $E$, the Schr\"odinger equation reads
\begin{eqnarray}
\left( t+g \right) \psi_{i-1} + \left( t-g \right) \psi_{i+1} &=& E \psi_{i}
	\label{seq: HN - bulk}
\end{eqnarray}
in the bulk and
\begin{equation}
\psi_{0} = \lim_{i \to \infty} \psi_{i} = 0
	\label{seq: HN - boundary}
\end{equation}
at the boundaries. We take an ansatz $\psi_{i} \sim \beta^{i}$ ($\beta \in \mathbb{C}$). From the bulk equation~(\ref{seq: HN - bulk}), we have
\begin{equation}
\frac{t+g}{\beta} + \left( t-g\right) \beta = E,
\quad\mathrm{i.e.,}\quad
\beta = \beta_{\pm} \left( E \right)
:= \frac{E \pm \sqrt{E^{2} - \left( t^{2} - g^{2} \right)}}{2 \left( t-g\right)}.
\end{equation}
Here, the absolute values of both $\left| \beta_{+} \left( E \right) \right|$ and $\left| \beta_{-} \left( E\right) \right|$ should be less than $1$ so that $\ket{E}$ can satisfy the semi-infinite boundary condition~(\ref{seq: HN - boundary}). To see this fact, suppose, for example, $\left| \beta_{+} \left( E \right) \right|$ is larger than $1$. We note that $\left| \beta_{+} \left( E \right) \right| = 1$ describes delocalized states, in which we are not interested here. Now, a generic eigenstate is given as $\psi_{i} = c_{+} \beta_{+}^{i} \left( E \right) + c_{-} \beta_{-}^{i} \left( E \right)$ with some coefficients $c_{+}, c_{-} \in \mathbb{C}$. However, because of $\left| \beta_{+} \left( E \right) \right| > 1$, the boundary condition given by Eq.~(\ref{seq: HN - boundary}) leads to $c_{+} + c_{-} = 0$, as well as $c_{+} = 0$ or $\beta_{+} = \beta_{-}$, either of which results in $\psi_{i} = 0$. Therefore, we need both $\left| \beta_{+} \left( E \right) \right| < 1$ and $\left| \beta_{-} \left( E \right) \right| < 1$ to obtain eigenstates, which in turn leads to
\begin{equation}
1 > \left| \beta_{+} \left( E \right) \beta_{-} \left( E \right) \right|
= \left| \frac{t+g}{t-g} \right|,
\quad\mathrm{i.e.,}\quad
g < 0.
	\label{seq: HN - ineq}
\end{equation}

Moreover, denoting $\beta =: \left| \beta \right| e^{\ii \varphi}$ ($\varphi \in \left[ 0, 2\pi \right]$), we have
\begin{equation}
E = \left[ \frac{t+g}{\left| \beta \right|} + \left( t-g \right) \left| \beta \right| \right] \cos \varphi + \ii \left[ -\frac{t+g}{\left| \beta \right|} + \left( t-g \right) \left| \beta \right| \right] \sin \varphi,
\end{equation}
which leads to
\begin{equation}
\frac{\left( \mathrm{Re}\,E\right)^{2}}{\left( 2t\right)^{2}} + \frac{\left( \mathrm{Im}\,E\right)^{2}}{\left( 2g\right)^{2}}
= \frac{1}{\left( 2t \right)^{2}} \left[ \frac{t+g}{\left| \beta \right|} + \left( t-g \right) \left| \beta \right| \right]^{2} \cos^{2} \varphi
+ \frac{1}{\left( 2g \right)^{2}} \left[ -\frac{t+g}{\left| \beta \right|} + \left( t-g \right) \left| \beta \right| \right]^{2} \sin^{2} \varphi.
\end{equation}
The right-hand side of this equation reaches a maximum for $\cos \varphi = 0$ or $\sin \varphi = 0$. Hence, we have
\begin{equation}
\frac{\left( \mathrm{Re}\,E\right)^{2}}{\left( 2t\right)^{2}} + \frac{\left( \mathrm{Im}\,E\right)^{2}}{\left( 2g\right)^{2}}
\leq \mathrm{max} \left\{ \frac{1}{\left( 2t \right)^{2}} \left[ \frac{t+g}{\left| \beta \right|} + \left( t-g \right) \left| \beta \right| \right]^{2},~\frac{1}{\left( 2g \right)^{2}} \left[ -\frac{t+g}{\left| \beta \right|} + \left( t-g \right) \left| \beta \right| \right]^{2} \right\}.
\end{equation}
Since both of the two in the right-hand side reach a maximum for $\left| \beta \right| = 1$ because of Eq.~(\ref{seq: HN - ineq}), we have 
\begin{equation}
\frac{\left( \mathrm{Re}\,E\right)^{2}}{\left( 2t\right)^{2}} + \frac{\left( \mathrm{Im}\,E\right)^{2}}{\left( 2g\right)^{2}}
\leq 1.
\end{equation}
Thus, the semi-infinite Hatano-Nelson model indeed has the infinite number of boundary states for $E \in \mathbb{C}$ with the nontrivial winding number $W \left( E \right) = \mathrm{sgn} \left( g\right) = -1$. Remarkably, the above discussions are not necessarily applicable in finite systems with open boundaries, in which a stricter boundary condition $\psi_{0} = \psi_{N+1} = 0$ is imposed instead of Eq.~(\ref{seq: HN - boundary}).

%%%%%%%%%%
\section{SIV.~Open and semi-infinite boundary conditions}

{\it A proof of Eq.~(5).\,---} We show Eq.~(5) in the main text, i.e.,
\begin{align}
\lim_{N \to \infty} \sigma \left( H_{\rm OBC} \right) \subset \sigma \left( H_{\rm SIBC} \right).
	\label{suppleinclusion}
\end{align}
For concreteness, we consider a matrix representation of $H_{\rm OBC}$, where $H_{\rm OBC}$ is an $mN\times mN$ matrix with the hopping range $ l < \infty$, the number $m$ of internal degrees of freedom, and the number $N$ of unit cells. In the same matrix representation, the corresponding semi-infinite matrix $H_{\rm SIBC}$ can be described by
\begin{align}
H_{\rm SIBC}=
\begin{pmatrix}
H_{\rm OBC}&B\\
C&D
\end{pmatrix},
	\label{Seq: HSIBC - matrix}
\end{align}
where $B, C, D$ are the semi-infinite matrices. In particular, $C$ has nonzero entries only at the right corner. Equation~(\ref{suppleinclusion}) follows if we have
\begin{align}
\sigma \left( H_{\rm OBC} \right) \subset \sigma_{\epsilon \left( N \right)} \left( H_{\rm SIBC} \right),
\quad \lim_{N\rightarrow\infty} \epsilon \left( N \right) = 0,
	\label{pseudo_inclusion}
\end{align}
where $\sigma_{\epsilon} \left( H \right)$ is the $\epsilon$-pseudospectrum~\cite{Trefethen} defined by
\begin{align}
\sigma_{\epsilon} \left( H \right) = 
\{E\in\mathbb{C}~|~\| \left( H-E \right) \bm{u} \| < \epsilon~ {\rm for~some}~\bm{u}~{\rm with}~ \| \bm{u} \|=1\}.
\label{Seq: pseudospetrum}
\end{align}
Below, we show Eq.~(\ref{pseudo_inclusion}).

Let us assume $E\in \sigma \left( H_{\rm OBC} \right)$. Then, we have the corresponding normalized right eigenstate $\bm{u}_{N}$ and left eigenstates $\bm{v}_{N}$, i.e.,
\begin{align}
H_{\rm OBC}\,\bm{u}_{N} = E\,\bm{u}_N,\quad
H^{T}_{\rm OBC}\,\bm{v}_{N} =E\,\bm{v}_{N}.
	\label{Seq: eigenequations}
\end{align}
In general, if $\bm{u}_{N}$ is localized at the left end, $\bm{v}_{N}$ is localized at the right end, and vice versa \cite{KSUS-19}. Accordingly, these eigenstates may behave in the following manner:
\begin{itemize}
\item[(i)] Both $\bm{u}_{N}$ and $\bm{v}_{N}$ are delocalized and extended into the bulk.
\item[(ii)] While $\bm{u}_{N}$ is localized only at one end, $\bm{v}_{N}$ is localized at the other end.
\item[(iii)] Both $\bm{u}_{N}$ and $\bm{v}_{N}$ are localized at both ends.
\end{itemize}  
In the case~(ii), we assume without loss of generality that $\bm{u}_{N}$ ($\bm{v}_{N}$) is localized only at the left (right) end. If $\bm{u}_{N}$ ($\bm{v}_{N}$) is localized only at the right (left) end, the same argument follows if we consider $\bm{v}_{N}$ and $H^{T}_{\rm OBC}$ instead of $\bm{u}_{N}$ and $H_{\rm OBC}$ because of $\sigma \left( A \right) = \sigma~( A^T )$. For example, the skin modes of the Hatano-Nelson model correspond to the case~(ii). The topologically protected zero modes in the finite Su-Schrieffer-Heeger model correspond to the case~(iii); they are localized at both ends in $H_{\rm OBC}$ with finite $N$, whereas each of them is localized only at one end in infinite systems. Besides, the skin modes of the reciprocal non-Hermitian model described by Eq.~(10) in the main text can correspond to the case~(iii).

We first consider the cases~(i) and (ii). Combining $\bm{u}_{N}$ and the infinite-dimensional zero vector $\left( 0~0~\cdots \right)$, we introduce the following infinite-dimensional vector $\bm{u}$ by
\begin{align}
\bm{u} := \left( \bm{u}_N~0~0~\cdots \right)^T.
\end{align}
From Eqs.~(\ref{Seq: HSIBC - matrix}) and (\ref{Seq: eigenequations}), we obtain
\begin{align}
\| \left( H_{\rm SIBC}-E \right) \bm{u} \| &= \|C \bm{u}_N \|.
	\label{normrelation}
\end{align}
Since the matrix $C$ has nonzero entries only at the upper right corner, it couples only to the right-end components of $\bm{u}_N$. In the case~(i), the normalization condition $\| \bm{u}_N \| = 1$ implies that these right-end components are $\mathcal{O}~( 1/\sqrt{N} )$. In the case (ii), on the other hand, they are exponentially small for $N$.  
Therefore, for $\epsilon \left( N \right) := \alpha \| C\bm{u}_N \|$ with a constant $\alpha>1$, we have $\lim_{N\rightarrow\infty}\epsilon \left( N \right) = 0$ and hence Eq.~(\ref{pseudo_inclusion}).

In the case~(iii), $\bm{u}_N$ is localized at both ends through a tunneling coupling between the left and right ends. This tunneling coupling also provides a nearly degenerate eigenstate $\bm{u}'_N$ of $H_{\rm OBC}$,
\begin{align}
H_{\rm OBC}\,\bm{u}'_N = \left( E + \delta_{N} \right) \bm{u}'_N,
\quad
\lim_{N\rightarrow \infty} \delta_{N} =0.
\end{align}
As a superposition of $\bm{u}_{N}$ and $\bm{u}'_{N}$, we take a normalized state $\bm{w}_{N} := a \bm{u}_{N} + b \bm{u}'_{N}$. Here, the coefficients $a, b \in \mathbb{C}$ are chosen such that $\bm{w}_{N}$ is localized only at the left end. Consequently, we have
\begin{align}
\left( H_{\rm OBC} - E \right) \bm{w}_{N} = b\,\delta_N\,\bm{u}'_{N},
\end{align} 
and hence $\bm{w}_{N}$ is no longer an eigenstate of $H_{\rm OBC}$. Nevertheless, introducing the infinite-dimensional vector $\bm{w}$ as
\begin{align}
\bm{w}:= \left( \bm{w}_N~0~0~\cdots \right)^T,
\end{align}
we have 
\begin{align}
\| \left( H_{\rm SIBC}-E \right) \bm{w} \| = \| C\bm{w}_{N} \| + \left| b \right| \left| \delta_{N} \right| \| \bm{u}'_{N} \|.
\end{align}
For $N\rightarrow \infty$, the right-hand side goes to zero since $\bm{w}_N$ decays near the right end and $\delta_{N} \to0$. Hence, we obtain Eq.~(\ref{pseudo_inclusion}) for $\epsilon \left( N \right) := \alpha \left( \| C\bm{w}_N \| + \left| b \right| \left| \delta_{N} \right| \| \bm{u}'_{N} \| \right)$ with a constant $\alpha > 0$.

\bigskip

{\it Another proof of Eq.~(5).\,---} We provide another proof of Eq.~(\ref{suppleinclusion}) with precise definitions of the spectra: 
\begin{align}
\sigma \left( H_{\rm SIBC} \right)&:=\{ E\in \mathbb{C}~|~\text{$H_{\rm SIBC} - E$ is not invertible}\}, \\
\lim_{N \to \infty} \sigma \left( H_{\rm OBC} \right)&:=\lim_{N\rightarrow\infty}\sup\,\sigma \left( H_{\rm OBC} \right),
\end{align}
where $\lim_{N\rightarrow\infty}\sup\,\sigma \left( H_{\rm OBC} \right)$ stands for the set of all $E \in \mathbb{C}$ for which there are $N_{1} < N_{2} < N_{3} < \cdots$ and $E_{N_{k}} \in \sigma\,( H_{\rm OBC})$ with $N=N_k$ such that $E_{N_{k}} \xrightarrow{k\to\infty} E$~\cite{Bottcher}. In other words, $E \in \lim_{N \to \infty} \sigma \left( H_{\rm OBC} \right)$ means that for sufficiently large $N_{k}$'s there exists a sequence of eigenvalues $E_{N_{k}}$ of $H_{\rm OBC}$ with $N=N_k$ that converges to $E$.

%a physically reasonable assumption on the spectra of Hermitian systems. 
To prove Eq.~(5), we show that 
$E\notin \lim_{N\rightarrow\infty}\sigma(H_{\rm OBC})$ if $E \notin \sigma \left( H_{\rm SIBC} \right)$.
For $E\notin \sigma\left(H_{\rm SIBC}\right)$,  we employ the following lemma (see its proofs at the end of this section),
\begin{itemize}
\item[] \textbf{Lemma}~~If $H_{\rm SIBC} - E$ is invertible, the extended Hermitian Hamiltonian
\begin{equation}
\tilde{H}_{\rm SIBC} \left( E \right) := \left( \begin{array}{@{\,}cc@{\,}} 
	0 & H_{\rm SIBC} - E \\
	H_{\rm SIBC}^{\dag} - E^{*} & 0 \\ 
	\end{array} \right)
\end{equation}
has a nonzero energy gap $2\Delta E > 0$.
\end{itemize}
First, we note that Lemma implies that the extended Hamiltonian $\tilde{H}_{\rm OBC}$ for the open boundary condition 
\begin{equation}
\tilde{H}_{\rm OBC} \left( E \right) := \left( \begin{array}{@{\,}cc@{\,}} 
	0 & H_{\rm OBC} - E \\
	( H_{\rm OBC} )^{\dag} - E^{*} & 0 \\ 
	\end{array} \right),
\end{equation}
also has a nonzero energy gap $2\Delta E>0$.
This result immediately follows from the reasonable assumption that a Hermitian Hamiltonian has a common energy gap between the semi-infinite boundary condition and the open boundary condition for a sufficiently large system size. 
Since a singular value of $H_{\rm OBC}- E$ is equal to a non-negative eigenvalue of $\tilde{H}_{\rm OBC} \left( E \right)$, 
this also indicates that the minimal singular value $s_{\rm min}$ of $H_{\rm OBC}-E$ satisfies $s_{\rm min}\ge \Delta E$.
Then, Weyl's inequality leads to
\begin{equation}
\left| \lambda_{\rm min} \right| \geq s_{\rm min} >\Delta E > 0,
\end{equation}
where $\lambda_{\rm min}$ is the eigenvalue of $H_{\rm OBC} - E$ with the smallest absolute value.
Therefore, there is no sequence of eigenvalues of $H_{\rm OBC}$ converging to $E$, that is, $E \notin \lim_{N \to \infty} \sigma \left( H_{\rm OBC} \right)$ 
(see also Lemma~{\bf 11.1} in Ref.~\cite{Bottcher} for the proof for Toeplitz operators).

Now we show Lemma in two different manners. 
Suppose that $H_{\rm SIBC}-E$ is invertible. 
Because Theorem~I in the main text implies that $\sigma \left( H \left( k \right)\right)\subset \sigma \left( H_{\rm SIBC} \right)$,  $H \left( k \right)-E$ is also invertible and thus $\det \left(H \left( k \right)-E \right) \neq 0$ for all $k \in \left[ 0, 2\pi \right]$. 
This means that the extended Hamiltonian $\tilde{H} \left( k, E \right)$, defined as
\begin{equation}
\tilde{H} \left( k,E \right) := \left( \begin{array}{@{\,}cc@{\,}} 
	0 & H \left( k \right) - E \\
	 H^{\dag} \left( k \right) - E^{*} & 0 \\ 
	\end{array} \right),
\end{equation}
has a nonzero energy gap if $H_{\rm SIBC}-E$ is invertible, because of $\det \tilde{H} \left( k,E \right) = \left| \det (H \left( k \right)-E) \right|^{2} \neq 0$. 
Now we use a reasonable assumption for a (one-dimensional) Hermitian Hamiltonian that the spectrum with the semi-infinite boundary consists of the bulk spectrum with the periodic boundary condition together with the (discrete) edge spectrum.
%
%As a result, the bulk real spectrum $\sigma\,( \tilde{H} \left( k \right) )$ of $\tilde{H}_{\rm SIBC}$ also has a nonzero energy gap 
%Now, for Hermitian Hamiltonians $\tilde{H}_{\rm SIBC}$ with the semi-infinite boundary, it is widely believed that when the bulk has %a nonzero energy gap, the spectrum $\sigma\,( \tilde{H}_{\rm SIBC} )$ is composed of the bulk spectrum $\sigma\,( \tilde{H} \left( k %\right))$ and the edge spectrum,
From this, we have 
\begin{equation}
\sigma\,( \tilde{H}_{\rm SIBC} \left( E \right) )= \sigma\,( \tilde{H} \left( k,E \right))~\cup~\{ \text{eigenvalues of (isolated) edge states of $\tilde{H}_{\rm SIBC} \left( E \right)$} \}.
\end{equation}
Since $\tilde{H} \left( k,E \right)$ has a gap as mentioned above, Lemma follows if $\tilde{H}_{\rm SIBC} \left( E \right)$ does not have zero-energy edge states.
The latter property can be shown by contraposition.
Suppose that there is a zero-energy edge state in $\tilde{H}_{\rm SIBC} \left( E \right)$. 
Thanks to chiral symmetry $\{\tilde{H}_{\rm SIBC} \left( E \right), \Gamma\} = 0$ with $\Gamma = \sigma_{z}$, the zero mode must be an eigenstate of $\Gamma$ with the eigenvalue (i.e., chirality) $\pm 1$. Then, the zero mode with positive chirality satisfies $\tilde{H}_{\rm SIBC} \left( E \right) \left( \ket{0}~0 \right)^{T} = 0$, which means $(H_{\rm SIBC}^{\dag}-E )\ket{0} = 0$ and thus $H_{\rm SIBC}-E$ is not invertible. Similarly, the zero mode with negative chirality satisfies $\tilde{H}_{\rm SIBC} \left( E \right) \left( 0~\ket{0} \right)^{T} = 0$, which means $(H_{\rm SIBC}-E) \ket{0} = 0$ and thus $H_{\rm SIBC}-E$ is not invertible. 
Therefore, by contraposition, $\tilde{H}_{\rm SIBC} \left( E \right)$ does not have zero-energy edge states if $H_{\rm SIBC}-E$ is invertible, leading to Lemma.

Another proof of Lemma is much simpler. If $H_{\rm SIBC}-E$ is invertible, there exists $\epsilon > 0$ such that $\| (H_{\rm SIBC}-E)^{-1}\| < \epsilon$ by definition. Since we have
\begin{align}
\| ( (H_{\rm SIBC}-E) (H_{\rm SIBC}-E)^{\dag} )^{-1} \| 
\leq \| (H_{\rm SIBC}-E)^{-1} \| \| ((H_{\rm SIBC}-E)^{\dag})^{-1} \|
< \epsilon^{2},
\end{align}
$(H_{\rm SIBC}-E)(H_{\rm SIBC}-E)^{\dag}$ is invertible. Similarly, $(H_{\rm SIBC}-E)^{\dag} (H_{\rm SIBC}-E)$ is invertible. Since we have $\sigma\,( \tilde{H}_{\rm SIBC}^2(E) ) = \sigma\,( (H_{\rm SIBC}-E)(H_{\rm SIBC}-E)^{\dag} ) \cup \sigma\,( (H_{\rm SIBC}-E)^{\dag} (H_{\rm SIBC}-E) )$,  $\tilde{H}_{\rm SIBC}(E)$ has a gap if $H_{\rm SIBC}-E$ is invertible.

\bigskip
{\it Converse of Eq.~(5).\,---} In a similar manner to the first proof of Eq.~(5), we can construct an approximate eigenstate of $H_{\rm OBC}$ from an eigenstate of $H_{\rm SIBC}$, which implies
\begin{equation}
\sigma \left( H_{\rm SIBC} \right) 
\subset
\sigma_{\epsilon \left( N \right)} \left( H_{\rm OBC} \right),
\quad \lim_{N\rightarrow\infty} \epsilon \left( N \right) = 0
    \label{Seq: inclusion-supset}
\end{equation}
for appropriate $\epsilon \left( N \right)$. We have $\lim_{N \to \infty} \sigma_{\epsilon \left( N \right)} \left( H_{\rm OBC} \right) = \lim_{N \to \infty} \sigma \left( H_{\rm OBC} \right)$ for Hermitian Hamiltonians or non-Hermitian Hamiltonians with certain symmetry, in which no skin effects occur. However, this is not the case for generic non-Hermitian Hamiltonians that can exhibit skin effects. Because of $\lim_{N\rightarrow\infty} \epsilon \left( N \right) = 0$, there exists $N$-independent $\epsilon_{0} > 0$ such that $\epsilon \left( N \right) < \epsilon_{0}$ for sufficiently large $N$, which satisfies $\sigma_{\epsilon \left( N \right)} \left( H_{\rm OBC} \right) \subset \sigma_{\epsilon_{0}} \left( H_{\rm OBC} \right)$. Now, let us take the $N \to \infty$ limit and the subsequent $\epsilon_{0} \to 0$ limit, yielding
\begin{equation}
    \lim_{N \to \infty} \sigma_{\epsilon \left( N \right)} \left( H_{\rm OBC} \right) \subset 
    \lim_{\epsilon_{0} \to 0} \lim_{N \to \infty} \sigma_{\epsilon_{0}} \left( H_{\rm OBC} \right).
\end{equation}
We notice the following fact (see its proof below):
\begin{equation}
   \lim_{\epsilon_{0} \to 0} \lim_{N \to \infty} \sigma_{\epsilon_{0}} \left( H_{\rm OBC} \right)
   = \sigma \left( H_{\rm SIBC} \right).
    \label{Seq: OBC-SIBC epsilon-N}
\end{equation}
The above discussion results in
\begin{equation}
    \sigma \left( H_{\rm SIBC} \right) \subset 
    \lim_{N \to \infty} \sigma_{\epsilon \left( N \right)} \left( H_{\rm OBC} \right) \subset
    \sigma \left( H_{\rm SIBC} \right),
\end{equation}
which implies
\begin{equation}
    \lim_{N \to \infty} \sigma_{\epsilon \left( N \right)} \left( H_{\rm OBC} \right) = \sigma \left( H_{\rm SIBC} \right).
\end{equation}
Therefore, $\sigma \left( H_{\rm SIBC} \right) \subset \lim_{N \to \infty} \sigma \left( H_{\rm OBC} \right)$ does not necessarily follow.

Finally, we prove Eq.~(\ref{Seq: OBC-SIBC epsilon-N}). Below, we assume that the norm in Eq.~(\ref{Seq: pseudospetrum}) is the standard Euclidean norm $\|\cdot\|_2$. 
First, we show $\lim_{\epsilon_{0} \to 0} \lim_{N \to \infty} \sigma_{\epsilon_{0}} \left( H_{\rm OBC} \right)\subset \sigma(H_{\rm SIBC})$.
Suppose $E \in \lim_{\epsilon_{0} \to 0} \lim_{N \to \infty} \sigma_{\epsilon_{0}} \left( H_{\rm OBC} \right)$. Then, because of $\lim_{\epsilon_{0} \to 0} \lim_{N \to \infty} \sigma_{\epsilon_{0}} \left( H_{\rm OBC} \right) \subset \lim_{N \to \infty} \sigma_{\epsilon} \left( H_{\rm OBC} \right)$ for any $\epsilon > 0$,
we have $E \in \lim_{N \to \infty} \sigma_{\epsilon} \left( H_{\rm OBC} \right)$,
 which 
implies $E\in \sigma_{\epsilon}(H_{\rm OBC})$ for sufficiently large $N$.
For the Euclidean norm and a finite-dimensional matrix $H$, the $\epsilon$-pseudospectrum in Eq.~(\ref{Seq: pseudospetrum}) is equivalently defined as~\cite{Trefethen}
\begin{align}
\sigma_{\epsilon} \left( H \right) = 
\{E\in\mathbb{C}~|~ s_{\rm min}\left( H-E \right) < \epsilon \},
\label{Seq: pseudospectrum2}
\end{align}
where $s_{\rm min} \left( H-E \right)$ is the smallest singular value of $H-E$, and thus we have 
$s_{\rm min} \left( H_{\rm OBC}-E \right) < \epsilon$ for sufficiently large $N$.
Since $s_{\rm min} \left( H_{\rm OBC}-E \right)$ is a minimal non-negative eigenvalue of $\tilde{H}_{\rm OBC} \left( E \right)$, 
this in turn leads to $0 \in\sigma_{\epsilon}\,( \tilde{H}_{\rm OBC} \left( E \right) )$ for any $\epsilon$ with sufficiently large $N$, 
that is, $0 \in \lim_{N \to \infty} \sigma\,(\tilde{H}_{\rm OBC} \left( E \right) )$. Meanwhile, we have
\begin{equation}
    \lim_{N \to \infty} \sigma\,(\tilde{H}_{\rm OBC} \left( E \right) ) = \sigma\,(\tilde{H}_{\rm SIBC} \left( E \right) ),
\end{equation}
since $\tilde{H}_{\rm OBC} \left( E \right)$ and $\tilde{H}_{\rm SIBC} \left( E \right)$ are Hermitian. Consequently, we have $0 \in \sigma\,(\tilde{H}_{\rm SIBC} \left( E \right) )$, which leads to $E \in \sigma(H_{\rm SIBC})$ from the contraposition of Lemma in this section. 
This establishes $\lim_{\epsilon_{0} \to 0} \lim_{N \to \infty} \sigma_{\epsilon_{0}} \left( H_{\rm OBC} \right)\subset \sigma(H_{\rm SIBC})$.
Conversely, if $E \in \sigma \left( H_{\rm SIBC} \right)$, $0 \in \sigma\,(\tilde H_{\rm SIBC} \left( E \right))$ and hence $0\in \lim_{N\rightarrow\infty}\sigma\,(\tilde H_{\rm OBC} \left( E \right))$.
Thus, $s_{\rm min} \left( H_{\rm OBC}-E \right)<\epsilon$ for any $\epsilon$ with sufficiently large $N$.   
From Eq.~(\ref{Seq: pseudospectrum2}), this means that $E\in \lim_{N\rightarrow\infty}\sigma_{\epsilon} \left( H_{\rm OBC} \right)$, which leads to
$E\in \lim_{\epsilon \to 0} \lim_{N \to \infty} \sigma_{\epsilon}\left( H_{\rm OBC} \right)$.
Therefore, we also have $\sigma \left( H_{\rm SIBC} \right)\subset \lim_{\epsilon \to 0} \lim_{N \to \infty} \sigma_{\epsilon} \left( H_{\rm OBC} \right)$, which completes a proof of Eq.~(\ref{Seq: OBC-SIBC epsilon-N}).

%%%%%%%%%%
\section{SV.~Trivial point-gap topology under the open boundary condition}

\begin{figure}[t]
\begin{center}
　　　\includegraphics[width=172mm, angle=0, clip]{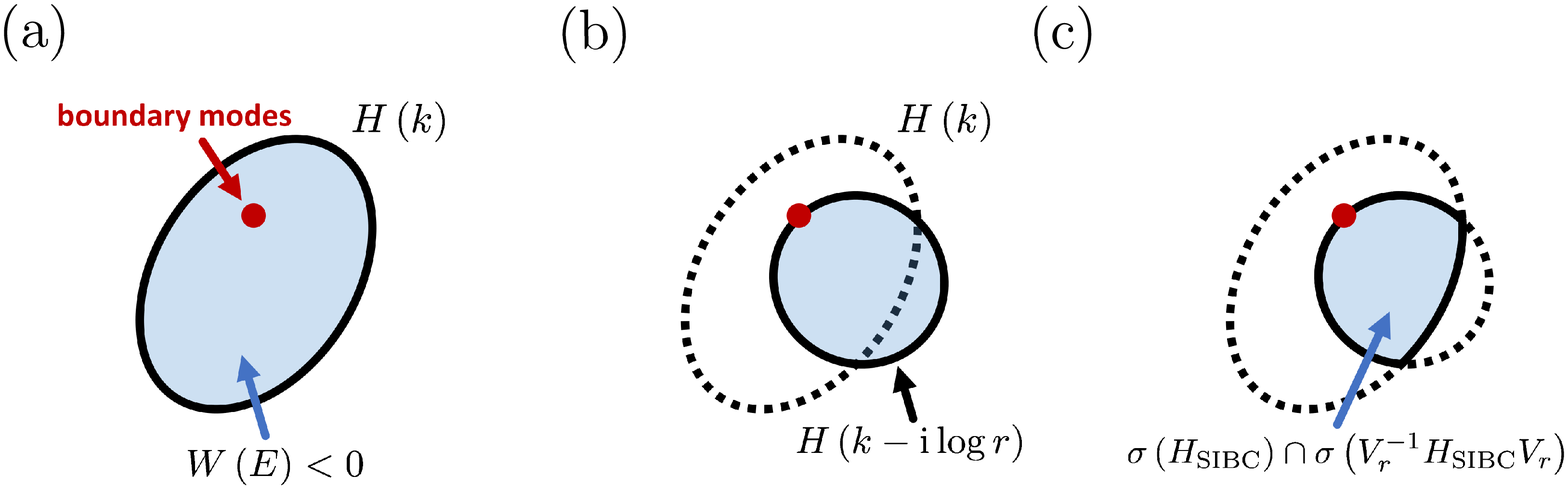}
　　　\caption{Intersection of $\sigma \left( H_{\rm SIBC} \right)$ and $\sigma \left( V_{r}^{-1} H_{\rm SIBC} V_{r} \right)$. (a)~When $H \left( k \right)$ has a point gap and $W \left( E \right) < 0$ for $E \in \mathbb{C}$, right boundary modes (red point) with eigenenergy $E$ appear in the semi-infinite system $H_{\rm SIBC}$ because of Theorem~I in the main text. The boundary modes are located inside the loop of $\sigma \left( H \left(k \right) \right)$ in the complex-energy plane. (b)~An appropriate imaginary gauge $V_{r}$ is chosen such that the boundary modes (red point) of $H_{\rm SIBC}$ are transformed to delocalized bulk modes of the transformed Hamiltonian $H \left( k-\ii 
\log r \right)$ with periodic boundaries. (c)~The intersection of $\sigma \left( H_{\rm SIBC} \right)$ and $\sigma \left( V_{r}^{-1} H_{\rm SIBC} V_{r} \right)$ is strictly smaller than $\sigma \left( H_{\rm SIBC} \right)$.}
　　　\label{fig: A7}
\end{center}
\end{figure}

We demonstrate
\begin{equation}
\sigma \left( H_{\rm SIBC} \right)\,\cap\,\sigma \left( V_{r}^{-1} H_{\rm SIBC} V_{r} \right)
~\subsetneq~ \sigma \left( H_{\rm SIBC} \right),\quad
r \neq 1,
\end{equation}
where $H_{\rm SIBC}$ is a semi-infinite system and $V_{r}^{-1} H_{\rm SIBC} V_{r}$ is the semi-infinite system obtained by the imaginary gauge transformation of $H_{\rm SIBC}$.
Here, $V_r$ is chosen so that some boundary modes of $H_{\rm SIBC}$ are transformed into delocalized modes.
First, when $H \left( k \right)$ has a point gap and $W \left( E \right) < 0$ for $E \in \mathbb{C}$, right boundary modes with eigenenergy $E$ appear in the corresponding semi-infinite system $H_{\rm SIBC}$ because of Theorem~I in the main text [Fig.~\ref{fig: A7}\,(a)]. Now, we perform an imaginary gauge transformation characterized by $V_{r}$ ($r \neq 1$) such that these boundary modes are transformed to delocalized bulk modes [Fig.~\ref{fig: A7}\,(b)]. Then, the original boundary modes of $H_{\rm SIBC}$ are transformed to delocalized eigenstates of the transformed periodic Hamiltonian $H \left( k - \ii \log r \right)$. In other words, the eigenenergy corresponding to these modes [the red point in Fig.~\ref{fig: A7}\,(b)] is on the edges of $\sigma \left( V_{r}^{-1} H_{\rm SIBC} V_{r} \right)$. Finally, we consider the intersection of $\sigma \left( H_{\rm SIBC} \right)$ and $\sigma \left( V_{r}^{-1} H_{\rm SIBC} V_{r} \right)$ [Fig.~\ref{fig: A7}\,(c)]. We trivially have $\sigma \left( H_{\rm SIBC} \right) \cap \sigma \left( V_{r}^{-1} H_{\rm SIBC} V_{r} \right) \subset \sigma \left( H_{\rm SIBC} \right)$; in addition, $\sigma \left( H_{\rm SIBC} \right) \cap \sigma \left( V_{r}^{-1} H_{\rm SIBC} V_{r} \right)$ cannot coincide with $\sigma \left( H_{\rm SIBC} \right)$. In fact, if they coincided with each other, it would contradict the fact that the original boundary modes are located inside the loop of $\sigma \left( H \left( k \right) \right)$. Thus, $\sigma \left( H_{\rm SIBC} \right) \cap \sigma \left( V_{r}^{-1} H_{\rm SIBC} V_{r} \right)$ is strictly smaller than $\sigma \left( H_{\rm SIBC} \right)$.

Next, we prove that $\bigcap_{r \in \left( 0, \infty \right)} \sigma \left( V^{-1}_{r} H_{\rm SIBC} V_{r} \right)$ in Eq.~(7) is a topologically trivial area (or open curves) of which interior satisfies $W \left( E \right)=0$. We prove it by contradiction. Let us first suppose $\bigcap_{r \in \left( 0, \infty \right)} \sigma \left( V^{-1}_{r} H_{\rm SIBC} V_{r} \right)$ includes a topologically nontrivial area $\mathbb{D}$ and focus on a complex energy $E$ inside $\mathbb{D}$. Then, there should exist $r \in \left( 0,\infty \right)$ such that $\sigma \left( V^{-1}_{r} H_{\rm SIBC} V_{r} \right)$ contains $E$ with $W \left( E \right) \neq 0$. Furthermore, Theorem~I implies that $V_r^{-1}H_{\rm SIBC} V_r$ hosts a boundary state with energy $E$. Now, we perform an imaginary gauge transformation $V_{r'}$ such that this boundary state is transformed into a delocalized bulk mode, which yields $E \in \sigma \left( V_{r'}^{-1} V_r^{-1} H_{\rm PBC} V_r V_{r'} \right)$. Since $E$ is defined to be inside $\mathbb{D}$, the intersection between $\mathbb{D}$ and $\sigma \left( V_{r'}^{-1} V_r^{-1} H_{\rm SIBC} V_r V_{r'} \right)$ is strictly smaller than $\mathbb{D}$. This contradicts the definition of $\mathbb{D} \subset \bigcap_{r \in \left( 0, \infty \right)} \sigma \left( V^{-1}_{r} H_{\rm SIBC} V_{r} \right)$.

%%%%%%%%%%
\section{SVI.~Spectral theory and non-Bloch band theory}

On the basis of spectral theory~\cite{Bottcher}, together with the arguments in Refs.~\cite{YW-18, Yokomizo-19}, we show
\begin{align}
\lim_{N\rightarrow\infty} \sigma \left( H_{\mathrm{OBC}} \right) = \bigcap_{r\in(0,\infty)} \sigma \left( V^{-1}_{r} H_{\mathrm{SIBC}} V_{r} \right).
	\label{Seq: spectrum OBC&SIBC}
\end{align}
We denote the hopping range of the system as $l < \infty$, and the number of internal degrees of freedom as $m$. We prove Eq.~(\ref{Seq: spectrum OBC&SIBC}) by contraposition, i.e., 
\begin{equation}
E\notin\lim_{N\rightarrow\infty}\sigma \left( H_{\mathrm{OBC}} \right)\quad\Leftrightarrow\quad
E\notin\bigcap_{r\in(0,\infty)}\sigma \left( V^{-1}_{r} H_{\mathrm{SIBC}} V_{r} \right).
	\label{Seq: contraposition}
\end{equation}
The non-Bloch band theory developed in Refs.~\cite{YW-18, Yokomizo-19} demonstrates that the spectrum of a non-Hermitian Hamiltonian $H_{\mathrm{OBC}}$ with open boundaries satisfies
\begin{align}
\lim_{N\rightarrow\infty} \sigma \left( H_{\mathrm{OBC}} \right) = 
\{E\in\mathbb{C}~|~|\beta_{lm}(E)|=|\beta_{lm+1}(E)|\},
	\label{Eq: YW-YM}
\end{align}
where $\beta_i (E)$'s ($i=1,\cdots, 2lm$) with $|\beta_1(E)|\leq\cdots \leq|\beta_{2lm}(E)|$ are the zeros of the $2lm$-th polynomial $\beta^{lm}\det \left( H (\beta)-E \right)$. In the following, we assume this result. Whereas a general argument in the main text only derives the inclusion relation of Eq.~(7), Eq.~(\ref{Eq: YW-YM}) leads to the stronger relation of Eq.~(\ref{Seq: spectrum OBC&SIBC}), as shown below.

We first notice
\begin{align}
E \notin \sigma \left( H_{\mathrm{SIBC}} \right)
\quad\Leftrightarrow\quad
W \left( E \right) =0,~{\rm and}~\det \left( H (\beta)-E \right)\neq0~\mathrm{for}~\beta \in \mathbb{T} := \{\beta\in\mathbb{C}~|~|\beta|=1\}
\end{align}
because of the index theorem (Theorem~I in the main text). Here, $\beta^{lm} \det \left( H (\beta)-E \right)$ is the $2lm$-th polynomial for $\beta$ such that
\begin{align}
\det \left( H (\beta)-E \right) = a_{-lm} \beta^{-lm} + a_{-lm+1} \beta^{-lm+1} + \cdots a_{lm} \beta^{lm},\quad
a_{i} \in \mathbb{C}.
\end{align}
Because of the argument principle, the winding number $W \left( E \right)$ for $H \left( \beta \right)$ with $\beta \in \mathbb{T}$ is given by the difference of the numbers of the zeros and the poles of $\det \left( H (\beta)-E \right)$ inside the disk $\mathbb{D} := \{ \beta\in\mathbb{C}~|~|\beta|<1\}$. Hence, the numbers of the zeros and the poles in $\mathbb{D}$ coincide with each other for $W \left( E \right) = 0$. In addition, $\beta=0$ is the only pole of $\det \left( H (\beta)-E \right)$ with the multiplicity $lm$. Thus, we have the following lemma:

\begin{itemize}
\item[] \textbf{Lemma}~~We have $E \notin \sigma \left( H_{\mathrm{SIBC}} \right)$ if and only if the $2lm$-th polynomial $\beta^{lm}\det \left( H (\beta)-E \right)$ has no zeros on $\mathbb{T}$ and has $lm$ zeros in $\mathbb{D}$. Similarly, we have $E \notin \sigma \left( V_{r}^{-1} H_{\mathrm{SIBC}} V_{r} \right)$ if and only if $\beta^{lm}\det \left( H (\beta)-E \right)$ has no zeros on $r^{-1}\mathbb{T} := \{\beta \in\mathbb{C}~|~|\beta|=r^{-1}\}$ and has $lm$ zeros in $r^{-1}\mathbb{D} := \{\beta \in\mathbb{C}~|~|\beta|<r^{-1}\}$.
\end{itemize}

Here, we note that Eq.~(5) in the main text immediately follows from the former part of Lemma. For $E \notin \sigma \left( H_{\rm SIBC} \right)$, Lemma leads to $\left| \beta_{lm} \left( E \right) \right| \neq \left| \beta_{lm + 1} \left( E \right) \right|$. Then, we have $E \notin \lim_{N \to \infty} \sigma \left( H_{\rm OBC} \right)$ because of Eq.~(\ref{Eq: YW-YM}), which implies Eq.~(5) in the main text.

Now, we show Eq.~(\ref{Seq: spectrum OBC&SIBC}). If we have $E \notin \lim_{N \to \infty} \sigma \left( H_{\rm OBC} \right)$, Eq.~(\ref{Eq: YW-YM}) implies that there exists $r_{\times} > 0$ such that
\begin{align}
|\beta_{lm} \left( E \right)| < r_{\times}^{-1} < | \beta_{lm+1} \left( E \right)|,
	\label{eq: rx}
\end{align}
and vice versa. Equation~(\ref{eq: rx}) implies that $\beta^{lm} \det \left( H \left( \beta \right) - E \right)$ has no zeros on $r_{\times}^{-1} \mathbb{T}$ and $lm$ zeros in $r_{\times}^{-1} \mathbb{D}$. Then, Eq.~(\ref{eq: rx}) also means $E \notin \sigma~( V_{r_{\times}}^{-1} H_{\rm SIBC} V_{r_{\times}} )$ and hence $E\notin\bigcap_{r\in(0,\infty)}\sigma \left( V^{-1}_{r} H_{\mathrm{SIBC}} V_{r} \right)$ because of Lemma, resulting in Eq.~(\ref{Seq: contraposition}). We thus have Eq.~(\ref{Seq: spectrum OBC&SIBC}).

%%%%%%%%%%
\section{SVII.~Non-Hermitian Su-Schrieffer-Heeger model}

\begin{figure}[b]
\begin{center}
　　　\includegraphics[width=172mm, angle=0, clip]{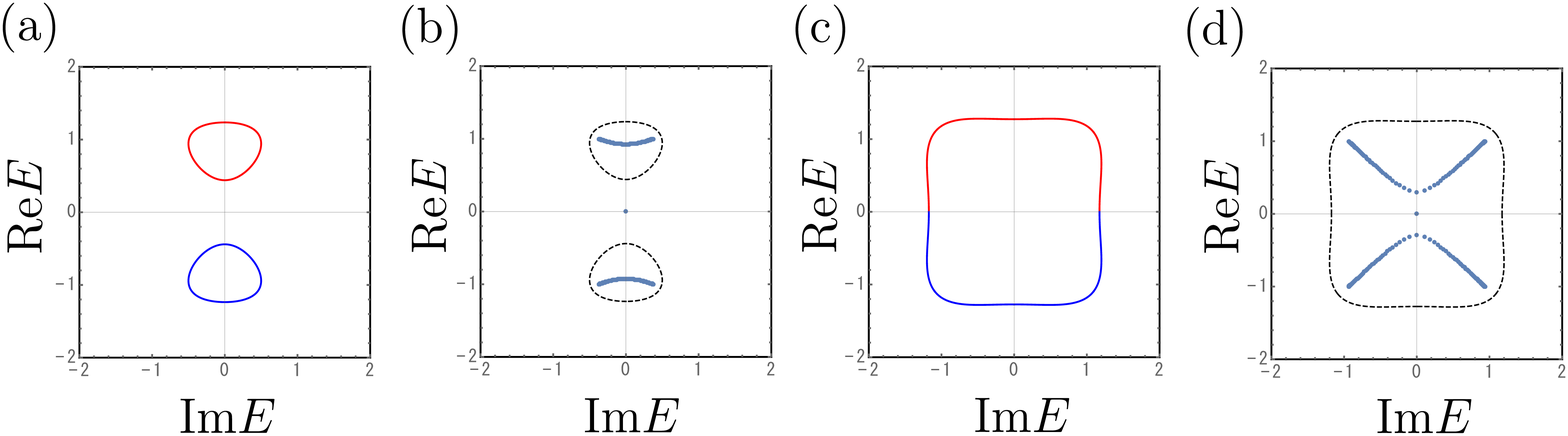}
　　　\caption{Non-Hermitian Su-Schrieffer-Heeger model with asymmetric hopping. (a, b)~Complex spectra under (a)~periodic and (b)~open boundary conditions ($N=100$) with $v=1/3, w=1, g=1/2$. (a)~Under the periodic boundary condition, each of the two bands (red and blue curves) forms a loop in the complex-energy plane. (b)~Under the open boundary condition, the energy bands (blue solid curves) form arcs and are different from those under the periodic boundary condition (black dotted curves), which signals the non-Hermitian skin effect. Still, a line gap for the imaginary axis is open (i.e., $\mathrm{Re}\,E \neq 0$) and the zero modes appear because of the line-gap topology. (c, d)~Complex spectra under (c)~periodic and (d)~open boundary conditions ($N=100$) with $v=3/4, w=1, g=6/5$.}
　　　\label{fig: SSH}
\end{center}
\end{figure}

As a prototypical non-Hermitian model that exhibits both skin effect and modified BBC, we investigate a non-Hermitian extension of the Su-Schrieffer-Heeger model~\cite{SSH-79} with asymmetric hopping~\cite{Lee-16, Kunst-18, YW-18, Yokomizo-19}:
\begin{equation}
H_{\rm SSH} \left( k \right)
= \left( v+w\cos k \right) \sigma_{x} + \left( w \sin k + \ii g \right) \sigma_{y}
= \left( \begin{array}{@{\,}cc@{\,}} 
	0 & v+g+we^{-\ii k} \\
	v-g+we^{\ii k} & 0 \\ 
	\end{array} \right),
\end{equation}
with $v, w, g \in \mathbb{R}$. In contrast with the Hatano-Nelson model, $H_{\rm SSH} \left( k \right)$ respects sublattice symmetry:
\begin{equation}
S H_{\rm SSH} \left( k \right) S^{-1}
= - H_{\rm SSH} \left( k \right),\quad
S^{2} =1,
	\label{eq: SSH - SLS}
\end{equation}
with $S := \sigma_{z}$. As a result of sublattice symmetry, $H_{\rm SSH} \left( k \right)$ possesses $\mathbb{Z}$ topological phases in the presence of a line gap~\cite{KSUS-19}. The corresponding topological invariant (winding number $W_{\rm L}$) has the same form as that in the Hermitian case,
\begin{eqnarray}
W_{\rm L}
= \oint_{\rm BZ} \frac{dk}{4\pi\ii} \mathrm{tr} \left[ S H^{-1}_{\rm SSH} \left( k \right) \frac{dH_{\rm SSH} \left( k \right)}{dk} \right].
\end{eqnarray}
On the other hand, $H_{\rm SSH}(k)$ possesses $\mathbb{Z} \oplus \mathbb{Z}$ topological phases in the presence of a point gap, i.e., two independent $\mathbb{Z}$ topological invariants are well defined~\cite{KSUS-19}. One of these invariants coincides with $W_{\rm L}$, and the other is the following $W_{\rm P}$, which is intrinsic to point-gap topology (see also 1d class A with $S$ in Table~\ref{tab:SFH_quotient_AZ_add}):
\begin{eqnarray}
W_{\rm P}
&=& \oint_{\rm BZ} \frac{dk}{2\pi\ii} \frac{d}{dk} \log \det H_{\rm SSH} \left( k \right).
\end{eqnarray}
For Hermitian systems, $W_{\rm P}=0$.
Notably, $W_{\rm L}$ can be a half integer in non-Hermitian systems, while it is always an integer when $W_{\rm P} = 0$. In the above, the reference line for the line gap is chosen as a line in the complex-energy plane that crosses $E = 0$, and the reference point for the point gap is chosen as $E = 0$, so that they will respect sublattice symmetry. Whereas they may be chosen as lines and points away from these symmetric lines and points, respectively, the consequent topological phases reduce to those without sublattice-symmetry protection.

As shown in Fig.~\ref{fig: SSH}, the non-Hermitian Su-Schrieffer-Heeger model exhibits the skin effect under the open boundary condition, which originates from the intrinsic point-gap topology as elucidated in the main text. Consistent with our discussions in the main text, each of the two energy bands forms an arc in the complex-energy plane. Notably, we have $W_{\rm P} = 0$ in terms of the reference point $E=0$ for Fig.~\ref{fig: SSH}\,(a, b). Nevertheless, we have a nontrivial winding number when we choose a reference point as an energy inside the loop determined by $H_{\rm SSH} \left( k \right)$, instead of the symmetric point $E = 0$ that respects sublattice symmetry. In this case, the topological classification reduces to that without symmetry despite sublattice symmetry of the Hamiltonian, as described above. On the other hand, we have $W_{\rm P} \neq 0$ for Fig.~\ref{fig: SSH}\,(c, d), which directly leads to the skin effect. 

In either case, the bulk spectrum under the open boundary condition has a line gap for the imaginary axis
(i.e., $\mathrm{Re}\,E \neq 0$). 
Therefore, $W_{\rm L}$ can be defined for the bulk Hamiltonian under the open boundary condition. 
According to the recipe in Ref.~\cite{Kunst-18, YW-18, Kunst-19, Yokomizo-19}, 
the open-boundary bulk Hamiltonian is given by 
\begin{equation}
H_{\rm SSH} \left( k - \ii \log r_{0} \right)
= \left( \begin{array}{@{\,}cc@{\,}} 
	0 & v+g+wr_{0}^{-1}e^{-\ii k} \\
	v-g+wr_{0}e^{\ii k} & 0 \\ 
	\end{array} \right),
\quad r_{0}:= \sqrt{\left| \frac{v-g}{v+g} \right|},
\end{equation}
which provides a nonzero $W_{\rm L}$ in either case of Fig.~\ref{fig: SSH}.
As a result, topologically protected zero modes emerge in Fig.~\ref{fig: SSH}\,(b,d).

We note in passing that sublattice symmetry in Eq.~(\ref{eq: SSH - SLS}) is distinct from chiral symmetry defined by~\cite{KSUS-19}
\begin{equation}
\Gamma H^{\dag} \left( k \right) \Gamma^{-1}
= - H \left( k \right),\quad
\Gamma^{2} =1,
	\label{eq: SSH - CS}
\end{equation}
where $\Gamma$ is a unitary operator. Whereas Eqs.~(\ref{eq: SSH - SLS}) and (\ref{eq: SSH - CS}) are equivalent to each other in Hermitian systems ($H^{\dag} \left( k \right) = H \left( k \right)$), they are not in non-Hermitian systems. Correspondingly, $H \left( k \right)$ with chiral symmetry possesses $\mathbb{Z}$ topological phases in the presence of a line gap for the real part of the spectrum, it does not possess nontrivial point-gap topology in terms of a reference point on the symmetric line (i.e., the imaginary axis)~\cite{KSUS-19}. For example, another non-Hermitian extension of the Su-Schrieffer-Heeger model~\cite{SSH-79} with balanced gain and loss~\cite{Esaki-11, Schomerus-13, Weimann-17-exp, St-Jean-17-exp} respects chiral symmetry instead of sublattice symmetry.

%%%%%%%%%%
\section{SVIII.~$\mathbb{Z}_{2}$ skin effect in two dimensions}

We discuss the non-Hermitian skin effects in higher dimensions. In particular, we investigate the following two-dimensional model
\begin{align}
H \left( \bm{k} \right) = \left( \sin k_x \right) \sigma_x + \left( \sin k_y \right) \sigma_y + \ii \Gamma \left(\cos k_x+\cos k_y-\mu \right)1_{2\times2},
	\label{seq: Z2 - 2D - Ham}
\end{align}
with $\Gamma\in\mathbb{C}$ and the $2 \times 2$ identity matrix $1_{2\times2}$. The energy dispersion is given as
\begin{align}
E_{\pm} \left( \bm{k} \right) = 
\pm\sqrt{\sin^2k_x+\sin^2 k_y}+\ii \Gamma \left( \cos k_x+\cos k_y-\mu \right).
\end{align}
Because $\mu$ represents a constant shift in the complex-energy plane, we set $\mu=0$ henceforth. Related models are investigated in Refs.~\cite{Zhou-19, Herviou-entanglement-19}.

This Hamiltonian respects time-reversal symmetry
\begin{equation}
T H^{T} \left( \bm{k} \right) T^{-1}
= H \left( -\bm{k} \right),\quad
T T^{*} = -1,
\end{equation}
with $T := \ii \sigma_{y}$, and hence belongs to symmetry class AII$^{\dagger}$. In the presence of a point gap for a reference point $E \in \mathbb{C}$, such systems possess topological phases characterized by the $\mathbb{Z}_{2}$ invariant
\begin{equation} \begin{split}
( -1 &)^{\nu \left( E \right)} := \prod_{\mathsf{X}=\mathrm{I, II}} \mathrm{sgn} \left\{
\frac{ \mathrm{Pf} \left[ \left( H \left( {\bm k}_{\mathsf{X}+} \right) - E \right) T \right] }{ \mathrm{Pf} \left[ \left( H \left( {\bm k}_{\mathsf{X}-} \right) - E \right) T \right] }
 \times \exp \left[ 
-\frac{1}{2} \int_{{\bm k} = {\bm k}_{\mathsf{X}-}}^{{\bm k} = {\bm k}_{\mathsf{X}+}} d \log \det \left[ \left( H \left( {\bm k} \right) - E \right) T \right]
\right] \right\},
\end{split} \end{equation}
where (${\bm k}_{\mathrm{I}+}, {\bm k}_{\mathrm{I}-}$) and (${\bm k}_{\mathrm{II}+}, {\bm k}_{\mathrm{II}-}$) are two pairs of time-reversal-invariant momenta~\cite{KSUS-19}. The Hamiltonian in Eq.~(\ref{seq: Z2 - 2D - Ham}) is topologically nontrivial (i.e., $\nu \left( E \right) = 1$ for some $E \in \mathbb{C}$) for $\Gamma \neq 0$.

\begin{figure}[t]
\begin{center}
　　　\includegraphics[width=172mm, angle=0, clip]{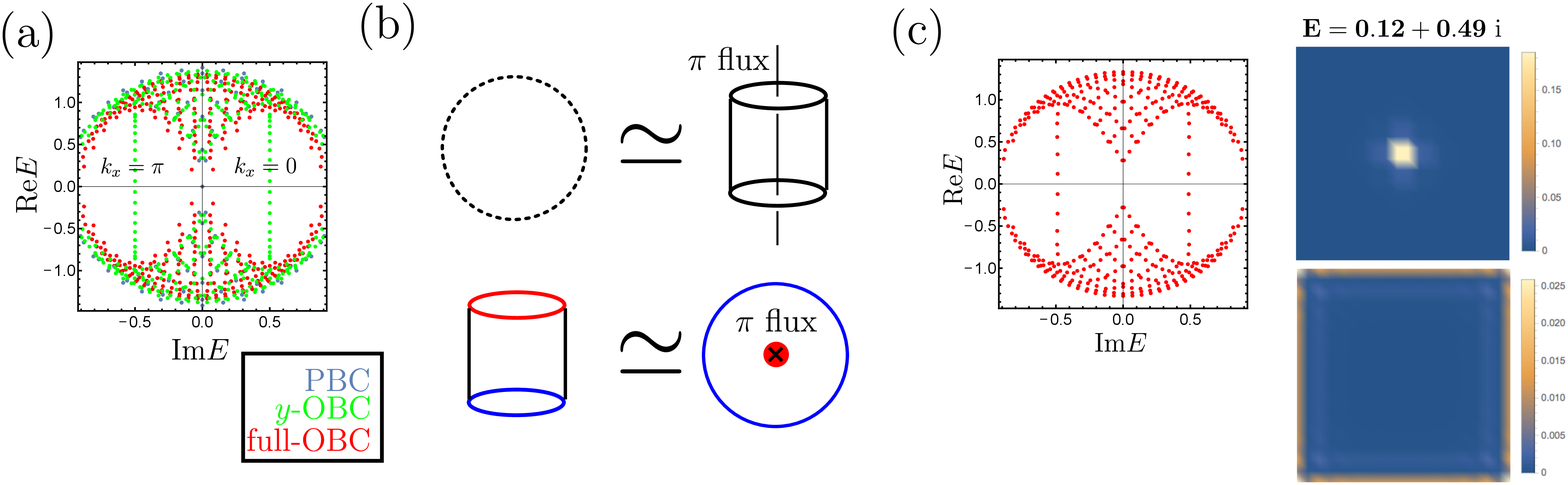}
　　　\caption{Symmetry-protected $\mathbb{Z}_{2}$ skin effect in two dimensions. All the calculations are performed with $\Gamma=0.5$ and $L=20$.  (a)~Complex energy spectra under the periodic boundary condition in both $x$ and $y$ directions (PBC, blue dots), the periodic boundary condition in the $x$ direction and the open boundary condition in the $y$ direction ($y$-OBC, green dots), and the open boundary condition in both $x$ and $y$ directions (full-OBC, red dots). (b)~Schematic picture of the equivalence between the full-OBC in the presence (absence) of a $\pi$ flux at the center and the $y$-OBC in the absence (presence) of a $\pi$ flux inside the cylinder. (c)~Complex energy spectrum and probability distributions of skin modes with $E = 0.12 + 0.49\ii$ in the full-OBC with a $\pi$ flux at the center. The skin mode localized at the defect in the center (upper) and that localized at the boundaries (lower) are degenerate.}
　　　\label{supp}
\end{center}
\end{figure}

In the following, we numerically and analytically investigate the corresponding Hamiltonian in real space under the following two boundary conditions:
\begin{itemize}
\item[(i)] Periodic boundary condition in the $x$ direction and open boundary condition in the $y$ direction ($y$-OBC).
\item[(ii)] Open boundary condition in both $x$ and $y$ directions (full-OBC).
\end{itemize} 
The case with the periodic boundary condition in the $y$ direction and the open boundary condition in the $x$ direction is equivalent to the case (i) in this model. The numerically obtained spectra for these boundary conditions are shown in Fig.~\ref{supp}\,(a). The case (i) shows the $\mathbb{Z}_2$ skin effect that has $2L\times2$ localized modes for $k_{x} = 0, \pi$ and no localized modes for $k_{x} \neq 0, \pi$, where the system size is $L \times L$. On the other hand, no skin effects occur for all the eigenmodes in the case (ii). These results do not change even in the presence of perturbations that preserve time-reversal symmetry. The result for the case (i) can be analytically explained, as follows. For simplicity, we take a unitary transformation such that $\sigma_y\rightarrow-\sigma_z$ while $\sigma_x$ is invariant. 

At $k_{x} = 0,\pi$, the tight-binding model of Eq.~(\ref{seq: Z2 - 2D - Ham}) in the $y$ direction is given by a stack of the two Hatano-Nelson models with spin-dependent asymmetric hopping:
\begin{align}
H_{\sigma_z=\pm1} = \sum_y \left( t_{\pm} c^\dagger_{y+1,\pm} c_{y,\pm} + t_{\mp}c^\dagger_{y,\pm}c_{y+1,\pm} \right),
	\label{stackedHatano}
\end{align}
with $t_\pm := \ii \left( \Gamma\pm1 \right)/2$. A similar spinful model is introduced in Ref.~\cite{Okuma-19}. Applying the imaginary gauge transformations in each spin sector, we have the Hermitian model
\begin{align}
H'_{\sigma_z=\pm1} = \sum_y \left( \sqrt{t_+t_-}c'^\dagger_{y+1,\pm}c'_{y,\pm}+\sqrt{t_+t_-}c'^\dagger_{y,\pm}c'_{y+1,\pm} \right).
\end{align}
Since these imaginary gauge transformations do not change the energy spectrum, it is given by 
\begin{equation}
\sigma \left( H_{y{\text{-OBC}}}\right) = \left\{ \sqrt{1-\Gamma^2} \cos \left( \frac{2\pi j}{L} \right) \pm \ii~\middle|~j=1,2,\cdots, L \right\}.
\end{equation} 
%This is nothing but the one-dimensional $\mathbb{Z}_2$ skin effect, and the
The corresponding wave functions are localized at both sides $y=1,L$ of the boundaries because of the Kramers degeneracy. In contrast to the one-dimensional model discussed in the main text, the present model does not include the coupling between the two spin sectors (such as the $\Delta$ term in the model in the main text). In a similar manner to the model in the main text, perturbations that preserve time-reversal symmetry cannot break the $\mathbb{Z}_{2}$ skin effect as long as the point gap is open. 

Furthermore, the skin modes are isolated from the bulk modes that exhibit no skin effect. For example, for $\Gamma=1$, the $2L\times2$ localized skin modes are degenerate at the two points $E=\pm \ii$, which are away from the other bulk modes. In fact, no skin effect occurs for $k_{x} \neq 0, \pi$ in this model. The $\left( \sin k_x \right) \sigma_x$ term with $k_{x} \neq 0, \pi$ behaves as a magnetic field that couples the two spin sectors in Eq.~(\ref{stackedHatano}). Reference~\cite{Okuma-19} shows that such a coupling leads to the suppression of the skin effect unless it is exponentially small with respect to $L$. Similarly, the $\left( \sin k_x \right) \sigma_x$ term in Eq.~(\ref{seq: Z2 - 2D - Ham}), which cannot be smaller than $\mathcal{O} \left( \pi/L \right)$ for $k_{x} \neq 0, \pi$, leads to the suppression of the skin effect. We note that if odd $L$ is chosen, $k_{x} = \pi$ is absent and thus the skin modes appear only for $k_{x}=0$.

Thus, only $\mathcal{O} \left( L \right)$ modes at $k_x=0,\pi$ from all the $\mathcal{O}~( L^{2} )$ modes exhibit the symmetry-protected skin effect in two dimensions. This can also be explained in terms of the extended Hermitian Hamiltonian. The eigenstate of the original non-Hermitian Hamiltonian can only be related with the exact zero mode of the extended Hermitian Hamiltonian, and the edge states of the extended Hamiltonian have nothing to do with the eigenstates of the original Hamiltonian, except for the zero modes. 

The suppression of skin effects for the case (ii) (full-OBC) can also be explained in terms of the absence of exact zero modes in the extended Hermitian Hamiltonian. In fact, the boundary eigenstates of the extended Hermitian Hamiltonian have at least $\mathcal{O} \left( \pi/L \right)$ energy and cannot be exact zero modes. This fact originates from the equivalence between a system with the full-OBC and the corresponding cylinder with a $\pi$ flux [Fig.~\ref{supp}\,(b)]. On the other hand, if we consider the system with the full-OBC and the additional $\pi$ flux, it is equivalent to the corresponding cylinder without fluxes [i.e, the case (i); Fig.~\ref{supp}\,(b)] and the skin effect occurs. Consistently, the $\mathbb{Z}_2$ skin effect indeed occurs in our model with such a $\pi$ flux, as shown in Fig.~\ref{supp}\,(c). Here, the $\pi$ flux is introduced by changing the sign of the hopping between $x=L/2$ and $x=L/2+1$ for $1 \leq y \leq L/2$. In addition, the skin modes localized at the boundaries and those localized at the defect in the center are degenerate in this model. Notably, the corresponding zero modes of the extended Hermitian Hamiltonian are a Kramers pair of Majorana zero modes localized at the $\pi$ flux. The presence of these defect zero modes is known to be a consequence of the strong topology of Hermitian topological superconductors with time-reversal symmetry in two dimensions~\cite{Qi-09}. In this sense, the presence of the skin modes localized at the $\pi$ flux is direct evidence of the strong point-gap topology in two dimensions.

In summary, the two-dimensional skin effects protected by time-reversal symmetry are characterized as follows:
\begin{itemize}
\item Under the open boundary condition in one direction and the periodic boundary condition in the other direction, only $\mathcal{O} \left( L \right)$ modes from all the $\mathcal{O}\,( L^{2} )$ modes exhibit the skin effect.
\item Under the open boundary condition in all the directions, no skin effects occur.
\item Even under the open boundary condition in all the directions, if we add a topological defect such as a $\pi$ flux in two dimensions, $\mathcal{O} \left( L \right)$ modes can exhibit the skin effect. The skin modes are localized at the boundaries or the defect.
\end{itemize}

%%%%%%%%%%
\section{SIX.~Homomorphism from line-gap to point-gap topology}

If a line gap is open, a point gap is also open with a reference point on the reference line. Hence, one can define a map from a line-gapped topological phase to a point-gaped one for each spatial dimension and symmetry class. Such maps are obtained as follows. 

Given a non-Hermitian Hamiltonian $H \left( \bk \right)$ in $d$ dimension, we introduce the extended Hermitian Hamiltonian~\cite{Gong-18, KSUS-19}
\begin{align}
    \tilde H \left( \bk \right) = \begin{pmatrix}
    0 & H \left( \bk \right) \\
    H^{\dag} \left( \bk \right) & 0 
    \end{pmatrix}_{\sigma},
\end{align}
which respects chiral symmetry $\Gamma H \left( {\bm k} \right) \Gamma^{-1} = -H \left( {\bm k} \right)$ with $\G := \s_z$ ($\sigma_i$ is the Pauli matrix in the extended space). Here, $\tilde{H} \left( {\bm k} \right)$ is gapped when $H \left( {\bm k} \right)$ has a point gap, and vice versa. Thus, topological classification for $H \left( {\bm k} \right)$ with respect to a point gap is nothing but the standard Hermitian one  for $\tilde{H} \left( {\bm k} \right)$~\cite{Gong-18, KSUS-19}. We denote the obtained topological phase as $K_{\rm P}$. On the other hand, when one keeps a line gap, an additional constraint arises on $\tilde{H} \left( {\bm k} \right)$. As shown in Ref.~\cite{KSUS-19}, a non-Hermitian Hamiltonian $H \left( \bk \right)$ with a real (an imaginary) line gap is continuously deformed to a Hermitian (an anti-Hermitian) Hamiltonian while keeping the line gap (and the point gap as well). Therefore, when one considers the classification of $\tilde{H} \left( {\bm k} \right)$ with keeping a real (an imaginary) line gap of $H \left( {\bm k} \right)$, $H \left( \bk \right)$ is supposed to be Hermitian (anti-Hermitian). Remarkably, Hermiticity (anti-Hermiticity) of $H \left( \bk \right)$ imposes one more additional chiral symmetry with $\G_{\rm r} = \s_y$ $(\G_{\rm i} = \s_x)$ on $\tilde H \left( \bk \right)$. Because of this additional symmetry, one has a different topological phase $K_{\rm L_r}$ ($K_{\rm L_i}$) with respect to a point gap in the presence of a real (an imaginary) line gap. Forgetting the additional chiral symmetry $\G_{\rm r}$ ($\G_{\rm i}$) defines a homomorphism $f_{\rm r}: K_{\rm L_{r}} \to K_{\rm P}$ ($f_{\rm i}: K_{\rm L_{i}} \to K_{\rm P}$) from $K_{\rm L_r}$ ($K_{\rm L_i}$) to $K_{\rm P}$. Thanks to the dimensional isomorphism of the $K$-theory, it suffices to compute $f_{\rm r}$ and $f_{\rm i}$ in zero dimension to get the homomorphisms in arbitrary spatial dimension. See Ref.~\cite{Shiozaki-19} for details. Tables~\ref{tab:SFH_AZ}, \ref{tab:SFH_AZdag}, and \ref{tab:SFH_AZ_add} summarize the homomorphisms $f_{\rm r}$ and $f_{\rm i}$ for 38-fold fundamental symmetry class of Ref.~\cite{KSUS-19}. See Ref.~\cite{KSUS-19} for the precise definition of each symmetry class.

If a point-gapped non-Hermitian Hamiltonian $H \left( \bk \right)$ lies in the image of either homomorphism $f_{\rm r}$ or $f_{\rm i}$, $H \left( \bk \right)$ can be continuously deformed to be Hermitian or anti-Hermitian, and thus the topological nature is attributed to the conventional Hermitian one. To the contrary, if $H \left( {\bm k} \right)$ does not, its topological nature is intrinsically non-Hermitian. In other words, the quotient group $K_{\rm P}/(\im f_{\rm r} \cup \im f_{\rm i})$, which is obtained from Tables~\ref{tab:SFH_AZ}, \ref{tab:SFH_AZdag}, and \ref{tab:SFH_AZ_add}, indicates the presence of the topological nature unique to non-Hermitian systems. A prime example of such intrinsic point-gap topology is found in one-dimensional systems without symmetry (class A), where the non-Hermitian skin effect occurs as a consequence of nontrivial topology as discussed in the main text. Tables~\ref{tab:SFH_quotient_AZ}, \ref{tab:SFH_quotient_AZdag}, and \ref{tab:SFH_quotient_AZ_add} summarize the quotient groups $K_{\rm P}/(\im f_{\rm r} \cup \im f_{\rm i})$ for all fundamental symmetry classes.

\begin{sidewaystable}
\caption{Homomorphisms $f_{\rm r}, f_{\rm i}$ from line-gap to point-gap topology for AZ class.}
	\label{tab:SFH_AZ}
\centering
{\scriptsize
$$
\begin{array}{ccccccccccccc}
\mbox{Symm. class} & \mbox{Gap} & d=0&d=1&d=2&d=3&d=4&d=5&d=6&d=7 \\
\hline \hline
{\rm A}& {\rm L} \to {\rm P} &\mathbb{Z} \to 0&0 \to \mathbb{Z}&\mathbb{Z} \to 0&0 \to \mathbb{Z}&\mathbb{Z} \to 0&0 \to \mathbb{Z}&\mathbb{Z} \to 0&0 \to \mathbb{Z}\\ 
&&&&&&&&&\\ 
\hline
{\rm AIII}& {\rm L_r}\to {\rm P}&0 \to \mathbb{Z}&\mathbb{Z} \to 0&0 \to \mathbb{Z}&\mathbb{Z} \to 0&0 \to \mathbb{Z}&\mathbb{Z} \to 0&0 \to \mathbb{Z}&\mathbb{Z} \to 0\\
&&&&&&&&&\\ 
&{\rm L_i} \to {\rm P}&\mathbb{Z}\oplus\mathbb{Z}\to\mathbb{Z}&0\to 0&\mathbb{Z}\oplus\mathbb{Z}\to\mathbb{Z}&0\to 0&\mathbb{Z}\oplus\mathbb{Z}\to\mathbb{Z}&0\to 0&\mathbb{Z}\oplus\mathbb{Z}\to\mathbb{Z}&0\to 0\\ 
&&(n,m)\mapsto n-m&&(n,m)\mapsto n-m&&(n,m)\mapsto n-m&&(n,m)\mapsto n-m&\\ 
\hline
{\rm AI}& {\rm L_r} \to {\rm P} &\mathbb{Z} \to \mathbb{Z}_2&0 \to \mathbb{Z}&0 \to 0&0 \to 0&2\mathbb{Z}\to 0&0\to 2\mathbb{Z}&\mathbb{Z}_2\to 0&\mathbb{Z}_2\to \mathbb{Z}_2 \\ 
&&n \mapsto n&&&&&&&n\mapsto n \\ 
&{\rm L_i} \to {\rm P}&\mathbb{Z}_2 \to \mathbb{Z}_2&\mathbb{Z}_2 \to \mathbb{Z}&\mathbb{Z} \to 0&0\to 0&0\to 0&0\to 2\mathbb{Z}&2\mathbb{Z}\to 0&0\to \mathbb{Z}_2\\ 
&&n \mapsto 0&&&&&&&\\
\hline
{\rm BDI}& {\rm L_r}\to {\rm P}&\mathbb{Z}_2 \to \mathbb{Z}_2&\mathbb{Z} \to \mathbb{Z}_2&0 \to \mathbb{Z}&0 \to 0&0 \to 0&2\mathbb{Z}\to 0&0\to 2\mathbb{Z}&\mathbb{Z}_2\to 0\\ 
&&n \mapsto n&n \mapsto n&&&&&&\\ 
&{\rm L_i \to P}&\mathbb{Z}_2\oplus\mathbb{Z}_2\to \mathbb{Z}_2&\mathbb{Z}_2\oplus \mathbb{Z}_2\to\mathbb{Z}_2&\mathbb{Z}\oplus \mathbb{Z}\to\mathbb{Z}&0\to 0&0\to 0&0\to 0&2\mathbb{Z}\oplus 2\mathbb{Z}\to 2\mathbb{Z}&0\to 0\\ 
&&(n,m)\mapsto n+m&(n,m)\mapsto n+m&(n,m)\mapsto n+m&&&&(n,m)\mapsto n+m&\\
\hline
{\rm D}& {\rm L} \to {\rm P}&\mathbb{Z}_2 \to 0&\mathbb{Z}_2 \to \mathbb{Z}_2&\mathbb{Z} \to \mathbb{Z}_2&0 \to \mathbb{Z}&0 \to 0&0 \to 0&2\mathbb{Z}\to 0&0\to 2\mathbb{Z}\\ 
&&&n \mapsto n&n \mapsto n&&&&&\\
\hline
{\rm DIII}& {\rm L_r}\to {\rm P}&0\to 2\mathbb{Z}&\mathbb{Z}_2 \to 0&\mathbb{Z}_2 \to \mathbb{Z}_2&\mathbb{Z} \to \mathbb{Z}_2&0 \to \mathbb{Z}&0 \to 0&0 \to 0&2\mathbb{Z}\to 0\\ 
&&&&n \mapsto n&n \mapsto n&&&&\\ 
&{\rm L_i \to P}&\mathbb{Z}\to 2\mathbb{Z}&0\to 0&\mathbb{Z}\to\mathbb{Z}_2&0\to\mathbb{Z}_2&\mathbb{Z}\to\mathbb{Z}&0\to 0&\mathbb{Z}\to 0&0\to 0\\ 
&&n\mapsto n&&n\mapsto n&&n\mapsto 2n&&\\
\hline
{\rm AII}& {\rm L_r} \to {\rm P}&2\mathbb{Z} \to 0&0\to 2\mathbb{Z}&\mathbb{Z}_2 \to 0&\mathbb{Z}_2 \to \mathbb{Z}_2&\mathbb{Z} \to \mathbb{Z}_2&0 \to \mathbb{Z}&0 \to 0&0 \to 0\\ 
&&&&&n \mapsto n&n \mapsto n&&&\\ 
&{\rm L_i} \to {\rm P}&0 \to 0&0 \to 2\mathbb{Z}&2\mathbb{Z} \to 0&0 \to \mathbb{Z}_2&\mathbb{Z}_2 \to \mathbb{Z}_2&\mathbb{Z}_2 \to \mathbb{Z}&\mathbb{Z} \to 0&0\to 0\\ 
&&&&&&n \mapsto 0&&&\\ 
\hline
{\rm CII}& {\rm L_r} \to {\rm P}&0 \to 0&2\mathbb{Z} \to 0&0\to 2\mathbb{Z}&\mathbb{Z}_2 \to 0&\mathbb{Z}_2 \to \mathbb{Z}_2&\mathbb{Z} \to \mathbb{Z}_2&0 \to \mathbb{Z}&0 \to 0\\ 
&&&&&&n \mapsto n&n \mapsto n&&\\ 
&{\rm L_i} \to {\rm P}&0\to 0&0\to 0&2\mathbb{Z}\oplus 2\mathbb{Z}\to 2\mathbb{Z}&0\to 0&\mathbb{Z}_2\oplus\mathbb{Z}_2\to \mathbb{Z}_2&\mathbb{Z}_2\oplus \mathbb{Z}_2\to\mathbb{Z}_2&\mathbb{Z}\oplus \mathbb{Z}\to\mathbb{Z}&0\to 0\\ 
&&&&(n,m)\mapsto n+m&&(n,m)\mapsto n+m&(n,m)\mapsto n+m&(n,m)\mapsto n+m\\
\hline
{\rm C}& {\rm L} \to {\rm P}&0 \to 0&0 \to 0&2\mathbb{Z} \to 0&0\to 2\mathbb{Z}&\mathbb{Z}_2 \to 0&\mathbb{Z}_2 \to \mathbb{Z}_2&\mathbb{Z} \to \mathbb{Z}_2&0 \to \mathbb{Z}\\ 
&&&&&&&n \mapsto n&n \mapsto n&\\
\hline
{\rm CI}& {\rm L_r} \to {\rm P}&0 \to \mathbb{Z}&0 \to 0&0 \to 0&2\mathbb{Z} \to 0&0\to 2\mathbb{Z}&\mathbb{Z}_2 \to 0&\mathbb{Z}_2 \to \mathbb{Z}_2&\mathbb{Z} \to \mathbb{Z}_2\\ 
&&&&&&&&n \mapsto n&n \mapsto n\\ 
& {\rm L_i} \to {\rm P}&\mathbb{Z}\to \mathbb{Z}&0\to 0&\mathbb{Z}\to 0&0\to 0&\mathbb{Z}\to 2\mathbb{Z}&0\to 0&\mathbb{Z}\to\mathbb{Z}_2&0\to\mathbb{Z}_2\\ 
&&n\mapsto 2n&&&&n\mapsto n&&n\mapsto n\\
\hline \hline
\end{array}
$$
}
\end{sidewaystable}

\begin{sidewaystable}
\caption{Homomorphisms $f_{\rm r}, f_{\rm i}$ from line-gap to point-gap topology for AZ$^\dag$ class. }
\label{tab:SFH_AZdag}
\centering
{\scriptsize
$$
\begin{array}{ccccccccccccc}
\mbox{Symm. class} & \mbox{Gap} & d=0&d=1&d=2&d=3&d=4&d=5&d=6&d=7 \\
\hline \hline
{\rm AI^\dag}& {\rm L} \to {\rm P}&\mathbb{Z} \to 0&0\to 0&0\to 0&0\to 2\mathbb{Z}&2\mathbb{Z}\to 0&0\to \mathbb{Z}_2&\mathbb{Z}_2\to \mathbb{Z}_2&\mathbb{Z}_2\to\mathbb{Z}\\ 
&&&&&&&&n\mapsto 0&\\
\hline
{\rm BDI^\dag}&{\rm L_r} \to {\rm P}&\mathbb{Z}_2 \to \mathbb{Z}&\mathbb{Z} \to 0&0\to 0&0\to 0&0\to 2\mathbb{Z}&2\mathbb{Z}\to 0&0\to \mathbb{Z}_2&\mathbb{Z}_2\to \mathbb{Z}_2\\ 
&&&&&&&&&n\mapsto 0\\ 
&{\rm L_i} \to {\rm P}&\mathbb{Z}\oplus \mathbb{Z}\to\mathbb{Z}&0\to 0&0\to 0&0\to 0&2\mathbb{Z}\oplus 2\mathbb{Z}\to 2\mathbb{Z}&0\to 0&\mathbb{Z}_2\oplus\mathbb{Z}_2\to\mathbb{Z}_2&\mathbb{Z}_2\oplus\mathbb{Z}_2\to\mathbb{Z}_2\\ 
&&(n,m)\mapsto n+m&&&&(n,m)\mapsto n+m&&(n,m)\mapsto n+m&(n,m)\mapsto n+m\\ 
\hline
{\rm D^\dag}&{\rm L_r} \to {\rm P}&\mathbb{Z}_2 \to \mathbb{Z}_2&\mathbb{Z}_2 \to \mathbb{Z}&\mathbb{Z} \to 0&0\to 0&0\to 0&0\to 2\mathbb{Z}&2\mathbb{Z}\to 0&0\to \mathbb{Z}_2\\ 
&&n \mapsto 0&&&&&&&\\
& {\rm L_i} \to {\rm P}&\mathbb{Z} \to \mathbb{Z}_2&0 \to \mathbb{Z}&0 \to 0&0 \to 0&2\mathbb{Z}\to 0&0\to 2\mathbb{Z}&\mathbb{Z}_2\to 0&\mathbb{Z}_2\to \mathbb{Z}_2 \\ 
&&n \mapsto n&&&&&&&n\mapsto n \\ 
\hline
{\rm DIII^\dag}& {\rm L_r} \to {\rm P}&0 \to \mathbb{Z}_2&\mathbb{Z}_2 \to \mathbb{Z}_2&\mathbb{Z}_2 \to \mathbb{Z}&\mathbb{Z} \to 0&0\to 0&0\to 0&0\to 2\mathbb{Z}&2\mathbb{Z}\to 0\\ 
&&&n \mapsto 0&&&&&&\\
&{\rm L_i} \to {\rm P}&\mathbb{Z}\to\mathbb{Z}_2&0\to\mathbb{Z}_2&\mathbb{Z}\to\mathbb{Z}&0\to 0&\mathbb{Z}\to 0&0\to 0&\mathbb{Z}\to 2\mathbb{Z}&0\to 0\\ 
&&n\mapsto n&&n\mapsto 2n&&&&n\mapsto n&\\ 
\hline
{\rm AII^\dag}& {\rm L} \to {\rm P}&2\mathbb{Z} \to 0&0 \to \mathbb{Z}_2&\mathbb{Z}_2 \to \mathbb{Z}_2&\mathbb{Z}_2 \to \mathbb{Z}&\mathbb{Z} \to 0&0\to 0&0\to 0&0\to 2\mathbb{Z}\\
&&&&n \mapsto 0&&&&&\\
\hline
{\rm CII^\dag}& {\rm L_r} \to {\rm P}&0 \to 2\mathbb{Z}&2\mathbb{Z} \to 0&0 \to \mathbb{Z}_2&\mathbb{Z}_2 \to \mathbb{Z}_2&\mathbb{Z}_2 \to \mathbb{Z}&\mathbb{Z} \to 0&0\to 0&0\to 0\\ 
&&&&&n \mapsto 0&&&&\\
&{\rm L_i} \to {\rm P}&2\mathbb{Z}\oplus 2\mathbb{Z}\to 2\mathbb{Z}&0\to 0&\mathbb{Z}_2\oplus\mathbb{Z}_2\to \mathbb{Z}_2&\mathbb{Z}_2\oplus \mathbb{Z}_2\to\mathbb{Z}_2&\mathbb{Z}\oplus \mathbb{Z}\to\mathbb{Z}&0\to 0&0\to 0&0\to 0\\ 
&&(n,m)\mapsto n+m&&(n,m)\mapsto n+m&(n,m)\mapsto n+m&(n,m)\mapsto n+m&&\\
\hline
{\rm C^\dag}& {\rm L_r} \to {\rm P}&0 \to 0&0 \to 2\mathbb{Z}&2\mathbb{Z} \to 0&0 \to \mathbb{Z}_2&\mathbb{Z}_2 \to \mathbb{Z}_2&\mathbb{Z}_2 \to \mathbb{Z}&\mathbb{Z} \to 0&0\to 0\\ 
&&&&&&n \mapsto 0&&&\\ 
& {\rm L_i} \to {\rm P}&2\mathbb{Z} \to 0&0\to 2\mathbb{Z}&\mathbb{Z}_2 \to 0&\mathbb{Z}_2 \to \mathbb{Z}_2&\mathbb{Z} \to \mathbb{Z}_2&0 \to \mathbb{Z}&0 \to 0&0 \to 0\\ 
&&&&&n \mapsto n&n \mapsto n&&&\\ 
\hline
{\rm CI^\dag}& {\rm L_r} \to {\rm P}&0 \to 0&0 \to 0&0 \to 2\mathbb{Z}&2\mathbb{Z} \to 0&0 \to \mathbb{Z}_2&\mathbb{Z}_2 \to \mathbb{Z}_2&\mathbb{Z}_2 \to \mathbb{Z}&\mathbb{Z} \to 0\\ 
&&&&&&&n \mapsto 0&&\\
& {\rm L_i} \to {\rm P}&\mathbb{Z}\to 0&0\to 0&\mathbb{Z}\to 2\mathbb{Z}&0\to 0&\mathbb{Z}\to\mathbb{Z}_2&0\to\mathbb{Z}_2&\mathbb{Z}\to\mathbb{Z}&0\to 0\\ 
&&&&n\mapsto n&&n\mapsto n&&n\mapsto 2n&\\
\hline \hline
\end{array}
$$
}
\end{sidewaystable}

\begin{sidewaystable}
\caption{Homomorphisms $f_{\rm r}, f_{\rm i}$ from line-gap to point-gap topology for AZ class with sublattice symmetry or pseudo-Hermiticity.}
\label{tab:SFH_AZ_add}
\centering
{\scriptsize
$$
\begin{array}{ccccccccccccc}
\mbox{AZ class} & \mbox{Add. symm.}& \mbox{Gap} &d=0&d=1&d=2&d=3&d=4&d=5&d=6&d=7 \\
\hline \hline
{\rm A}&\eta&{\rm L_r} \to {\rm P}&\mathbb{Z}\oplus\mathbb{Z}\to\mathbb{Z}&0\to 0&\mathbb{Z}\oplus\mathbb{Z}\to\mathbb{Z}&0\to 0&\mathbb{Z}\oplus\mathbb{Z}\to\mathbb{Z}&0\to 0&\mathbb{Z}\oplus\mathbb{Z}\to\mathbb{Z}&0\to 0\\
&&&(n,m)\mapsto n-m&&(n,m)\mapsto n-m&&(n,m)\mapsto n-m&&(n,m)\mapsto n-m&\\ 
&& {\rm L_i} \to {\rm P}&0 \to \mathbb{Z}&\mathbb{Z} \to 0&0 \to \mathbb{Z}&\mathbb{Z} \to 0&0 \to \mathbb{Z}&\mathbb{Z} \to 0&0 \to \mathbb{Z}&\mathbb{Z} \to 0\\
&&&&&&&&&&&\\ 
\hline
{\rm AIII}&S_+,\eta_{+}&{\rm L_r,L_i}\to {\rm P}&0\to 0&\mathbb{Z}\oplus\mathbb{Z}\to\mathbb{Z}&0\to 0&\mathbb{Z}\oplus\mathbb{Z}\to\mathbb{Z}&0\to 0&\mathbb{Z}\oplus\mathbb{Z}\to\mathbb{Z}&0\to 0&\mathbb{Z}\oplus\mathbb{Z}\to\mathbb{Z}\\ 
&&&&(n,m)\mapsto n-m&&(n,m)\mapsto n-m&&(n,m)\mapsto n-m&&(n,m)\mapsto n-m\\ 
\hline
{\rm AI}&\eta_+&{\rm L_r} \to {\rm P}&\mathbb{Z}\oplus \mathbb{Z}\to\mathbb{Z}&0\to 0&0\to 0&0\to 0&2\mathbb{Z}\oplus 2\mathbb{Z}\to 2\mathbb{Z}&0\to 0&\mathbb{Z}_2\oplus\mathbb{Z}_2\to\mathbb{Z}_2&\mathbb{Z}_2\oplus\mathbb{Z}_2\to\mathbb{Z}_2\\ 
&&&(n,m)\mapsto n+m&&&&(n,m)\mapsto n+m&&(n,m)\mapsto n+m&(n,m)\mapsto n+m\\ 
&&{\rm L_i} \to {\rm P}&\mathbb{Z}_2 \to \mathbb{Z}&\mathbb{Z} \to 0&0\to 0&0\to 0&0\to 2\mathbb{Z}&2\mathbb{Z}\to 0&0\to \mathbb{Z}_2&\mathbb{Z}_2\to \mathbb{Z}_2\\ 
&&&&&&&&&&n\mapsto 0\\ 
\hline
{\rm BDI}&S_{++},\eta_{++}&{\rm L_r,L_i}\to {\rm P}&\mathbb{Z}_2\oplus \mathbb{Z}_2\to\mathbb{Z}_2&\mathbb{Z}\oplus \mathbb{Z}\to\mathbb{Z}&0\to 0&0\to 0&0\to 0&2\mathbb{Z}\oplus 2\mathbb{Z}\to 2\mathbb{Z}&0\to 0&\mathbb{Z}_2\oplus\mathbb{Z}_2\to\mathbb{Z}_2\\ 
&&&(n,m)\mapsto n+m&(n,m)\mapsto n+m&&&&(n,m)\mapsto n+m&&(n,m)\mapsto n+m\\ 
\hline
{\rm D}&\eta_+& {\rm L_r} \to {\rm P}&\mathbb{Z}_2\oplus\mathbb{Z}_2\to \mathbb{Z}_2&\mathbb{Z}_2\oplus \mathbb{Z}_2\to\mathbb{Z}_2&\mathbb{Z}\oplus \mathbb{Z}\to\mathbb{Z}&0\to 0&0\to 0&0\to 0&2\mathbb{Z}\oplus 2\mathbb{Z}\to 2\mathbb{Z}&0\to 0\\ 
&&&(n,m)\mapsto n+m&(n,m)\mapsto n+m&(n,m)\mapsto n+m&&&&(n,m)\mapsto n+m&\\
&& {\rm L_i\to P}&\mathbb{Z}_2 \to \mathbb{Z}_2&\mathbb{Z} \to \mathbb{Z}_2&0 \to \mathbb{Z}&0 \to 0&0 \to 0&2\mathbb{Z}\to 0&0\to 2\mathbb{Z}&\mathbb{Z}_2\to 0\\ 
&&&n \mapsto n&n \mapsto n&&&&&&\\ 
\hline
{\rm DIII}&S_{--},\eta_{++}& {\rm L_r}\to {\rm P}&0\to 0&\mathbb{Z}_2\oplus\mathbb{Z}_2\to \mathbb{Z}_2&\mathbb{Z}_2\oplus \mathbb{Z}_2\to\mathbb{Z}_2&\mathbb{Z}\oplus \mathbb{Z}\to\mathbb{Z}&0\to 0&0\to 0&0\to 0&2\mathbb{Z}\oplus 2\mathbb{Z}\to 2\mathbb{Z}\\ 
&&&&(n,m)\mapsto n+m&(n,m)\mapsto n+m&(n,m)\mapsto n+m&&&&(n,m)\mapsto n+m\\ 
&&{\rm L_i}\to {\rm P}&0\to 0&\mathbb{Z}\to\mathbb{Z}_2&0\to\mathbb{Z}_2&\mathbb{Z}\to\mathbb{Z}&0\to 0&\mathbb{Z}\to 0&0\to 0&\mathbb{Z}\to 2\mathbb{Z}\\ 
&&&&n\mapsto n&&n\mapsto 2n&&&&n\mapsto n\\ 
\hline
{\rm AII}&\eta_+& {\rm L_r} \to {\rm P}&2\mathbb{Z}\oplus 2\mathbb{Z}\to 2\mathbb{Z}&0\to 0&\mathbb{Z}_2\oplus\mathbb{Z}_2\to \mathbb{Z}_2&\mathbb{Z}_2\oplus \mathbb{Z}_2\to\mathbb{Z}_2&\mathbb{Z}\oplus \mathbb{Z}\to\mathbb{Z}&0\to 0&0\to 0&0\to 0\\ 
&&&(n,m)\mapsto n+m&&(n,m)\mapsto n+m&(n,m)\mapsto n+m&(n,m)\mapsto n+m&&\\
&&{\rm L_i} \to {\rm P}&0 \to 2\mathbb{Z}&2\mathbb{Z} \to 0&0 \to \mathbb{Z}_2&\mathbb{Z}_2 \to \mathbb{Z}_2&\mathbb{Z}_2 \to \mathbb{Z}&\mathbb{Z} \to 0&0\to 0&0\to 0\\ 
&&&&&&n \mapsto 0&&&&\\
\hline
{\rm CII}&S_{++},\eta_{++}& {\rm L_r,L_i} \to {\rm P}&0\to 0&2\mathbb{Z}\oplus 2\mathbb{Z}\to 2\mathbb{Z}&0\to 0&\mathbb{Z}_2\oplus\mathbb{Z}_2\to \mathbb{Z}_2&\mathbb{Z}_2\oplus \mathbb{Z}_2\to\mathbb{Z}_2&\mathbb{Z}\oplus \mathbb{Z}\to\mathbb{Z}&0\to 0&0\to 0\\ 
&&&&(n,m)\mapsto n+m&&(n,m)\mapsto n+m&(n,m)\mapsto n+m&(n,m)\mapsto n+m&\\ 
\hline
{\rm C}&\eta_+& {\rm L_r} \to {\rm P}&0\to 0&0\to 0&2\mathbb{Z}\oplus 2\mathbb{Z}\to 2\mathbb{Z}&0\to 0&\mathbb{Z}_2\oplus\mathbb{Z}_2\to \mathbb{Z}_2&\mathbb{Z}_2\oplus \mathbb{Z}_2\to\mathbb{Z}_2&\mathbb{Z}\oplus \mathbb{Z}\to\mathbb{Z}&0\to 0\\ 
&&&&&(n,m)\mapsto n+m&&(n,m)\mapsto n+m&(n,m)\mapsto n+m&(n,m)\mapsto n+m\\
&& {\rm L_i} \to {\rm P}&0 \to 0&2\mathbb{Z} \to 0&0\to 2\mathbb{Z}&\mathbb{Z}_2 \to 0&\mathbb{Z}_2 \to \mathbb{Z}_2&\mathbb{Z} \to \mathbb{Z}_2&0 \to \mathbb{Z}&0 \to 0\\ 
&&&&&&&n \mapsto n&n \mapsto n&&\\ 
\hline
{\rm CI}&S_{--},\eta_{++}& {\rm L_r} \to {\rm P}&0\to 0&0\to 0&0\to 0&2\mathbb{Z}\oplus 2\mathbb{Z}\to 2\mathbb{Z}&0\to 0&\mathbb{Z}_2\oplus\mathbb{Z}_2\to \mathbb{Z}_2&\mathbb{Z}_2\oplus \mathbb{Z}_2\to\mathbb{Z}_2&\mathbb{Z}\oplus \mathbb{Z}\to\mathbb{Z}\\ 
&&&&&&(n,m)\mapsto n+m&&(n,m)\mapsto n+m&(n,m)\mapsto n+m&(n,m)\mapsto n+m\\ 
&& {\rm L_i} \to {\rm P}&0\to 0&\mathbb{Z}\to 0&0\to 0&\mathbb{Z}\to 2\mathbb{Z}&0\to 0&\mathbb{Z}\to\mathbb{Z}_2&0\to\mathbb{Z}_2&\mathbb{Z}\to\mathbb{Z}\\ 
&&&&&&n\mapsto n&&n\mapsto n&&n\mapsto 2n\\ 
\hline
\hline
\end{array}
$$
}
\end{sidewaystable}

\begin{sidewaystable}
\caption*{(Continued)}
\centering
{\scriptsize
$$
\begin{array}{ccccccccccccc}
\mbox{AZ class} & \mbox{Add. symm.}& \mbox{Gap} &d=0&d=1&d=2&d=3&d=4&d=5&d=6&d=7 \\
\hline \hline
{\rm A}&S&{\rm L} \to {\rm P}&0\to 0&\mathbb{Z}\to\mathbb{Z}\oplus\mathbb{Z}&0\to 0&\mathbb{Z}\to\mathbb{Z}\oplus\mathbb{Z}&0\to 0&\mathbb{Z}\to\mathbb{Z}\oplus\mathbb{Z}&0\to 0&\mathbb{Z}\to\mathbb{Z}\oplus\mathbb{Z}\\ 
&&&&n\mapsto(n,n)&&n\mapsto(n,n)&&n\mapsto(n,n)&&n\mapsto(n,n)\\ 
\hline
{\rm AIII}&S_-,\eta_-&{\rm L_r\to P}&\mathbb{Z}\to\mathbb{Z}\oplus\mathbb{Z}&0\to 0&\mathbb{Z}\to\mathbb{Z}\oplus\mathbb{Z}&0\to 0&\mathbb{Z}\to\mathbb{Z}\oplus\mathbb{Z}&0\to 0&\mathbb{Z}\to\mathbb{Z}\oplus\mathbb{Z}&0\to 0\\
&&&n\mapsto(n,n)&&n\mapsto(n,n)&&n\mapsto(n,n)&&n\mapsto(n,n)&\\
&&{\rm L_i\to P}&\mathbb{Z}\to\mathbb{Z}\oplus\mathbb{Z}&0\to 0&\mathbb{Z}\to\mathbb{Z}\oplus\mathbb{Z}&0\to 0&\mathbb{Z}\to\mathbb{Z}\oplus\mathbb{Z}&0\to 0&\mathbb{Z}\to\mathbb{Z}\oplus\mathbb{Z}&0\to 0\\
&&&n\mapsto(n,-n)&&n\mapsto(n,-n)&&n\mapsto(n,-n)&&n\mapsto(n,-n)&\\
\hline
{\rm AI}&S_-&{\rm L_r \to \rm P}&0\to 0&0\to\mathbb{Z}&0\to 0&2\mathbb{Z}\to\mathbb{Z}&0\to 0&\mathbb{Z}_2\to\mathbb{Z}&\mathbb{Z}_2\to 0&\mathbb{Z}\to\mathbb{Z}\\
&&&&&&n\mapsto 2n&&&&n\mapsto n\\
&&{\rm L_i \to \rm P}&0\to 0&\mathbb{Z}_2\to \mathbb{Z}&\mathbb{Z}_2\to 0&\mathbb{Z}\to \mathbb{Z}&0\to 0&0\to\mathbb{Z}&0\to 0&2\mathbb{Z}\to\mathbb{Z}\\
&&&&&&n\mapsto n&&&&n\mapsto 2n\\
\hline
{\rm BDI}&S_{-+},\eta_{+-}&{\rm L_r \to \rm P}&\mathbb{Z}\to \mathbb{Z}&0\to 0&0\to\mathbb{Z}&0\to 0&2\mathbb{Z}\to\mathbb{Z}&0\to 0&\mathbb{Z}_2\to\mathbb{Z}&\mathbb{Z}_2\to 0\\
&&&n\mapsto n&&&&n\mapsto 2n&&&\\
&&{\rm L_i \to \rm P}&\mathbb{Z}_2\to \mathbb{Z}&\mathbb{Z}_2\to 0&\mathbb{Z}\to \mathbb{Z}&0\to 0&0\to\mathbb{Z}&0\to 0&2\mathbb{Z}\to\mathbb{Z}&0\to 0\\
&&&&&n\mapsto n&&&&n\mapsto 2n&\\
\hline
{\rm D}&S_+&{\rm L \to \rm P}&\mathbb{Z}_2\to 0&\mathbb{Z}\to \mathbb{Z}&0\to 0&0\to\mathbb{Z}&0\to 0&2\mathbb{Z}\to\mathbb{Z}&0\to 0&\mathbb{Z}_2\to\mathbb{Z}\\
&&&&n\mapsto n&&&&n\mapsto 2n&&\\
\hline
{\rm DIII}&S_{-+},\eta_{-+}&{\rm L_r \to \rm P}&\mathbb{Z}_2\to \mathbb{Z}&\mathbb{Z}_2\to 0&\mathbb{Z}\to \mathbb{Z}&0\to 0&0\to\mathbb{Z}&0\to 0&2\mathbb{Z}\to\mathbb{Z}&0\to 0\\
&&&&&n\mapsto n&&&&n\mapsto 2n&\\
&&{\rm L_i \to \rm P}&\mathbb{Z}\to \mathbb{Z}&0\to 0&0\to\mathbb{Z}&0\to 0&2\mathbb{Z}\to\mathbb{Z}&0\to 0&\mathbb{Z}_2\to\mathbb{Z}&\mathbb{Z}_2\to 0\\
&&&n\mapsto n&&&&n\mapsto 2n&&&\\
\hline
{\rm AII}&S_-&{\rm L_r \to \rm P}&0\to 0&\mathbb{Z}_2\to \mathbb{Z}&\mathbb{Z}_2\to 0&\mathbb{Z}\to \mathbb{Z}&0\to 0&0\to\mathbb{Z}&0\to 0&2\mathbb{Z}\to\mathbb{Z}\\
&&&&&&n\mapsto n&&&&n\mapsto 2n\\
&&{\rm L_i \to \rm P}&0\to 0&0\to\mathbb{Z}&0\to 0&2\mathbb{Z}\to\mathbb{Z}&0\to 0&\mathbb{Z}_2\to\mathbb{Z}&\mathbb{Z}_2\to 0&\mathbb{Z}\to\mathbb{Z}\\
&&&&&&n\mapsto 2n&&&&n\mapsto n\\
\hline
{\rm CII}&S_{-+},\eta_{+-}&{\rm L_r \to \rm P}&2\mathbb{Z}\to \mathbb{Z}&0\to 0&\mathbb{Z}_2\to \mathbb{Z}&\mathbb{Z}_2\to 0&\mathbb{Z}\to \mathbb{Z}&0\to 0&0\to\mathbb{Z}&0\to 0\\
&&&n\mapsto 2n&&&&n\mapsto n&&&\\
&&{\rm L_i \to \rm P}&0\to \mathbb{Z}&0\to 0&2\mathbb{Z}\to \mathbb{Z}&0\to 0&\mathbb{Z}_2\to \mathbb{Z}&\mathbb{Z}_2\to 0&\mathbb{Z}\to \mathbb{Z}&0\to 0\\
&&&&&n\mapsto 2n&&&&n\mapsto n&\\
\hline
{\rm C}&S_+&{\rm L \to \rm P}&0\to 0&2\mathbb{Z}\to \mathbb{Z}&0\to 0&\mathbb{Z}_2\to \mathbb{Z}&\mathbb{Z}_2\to 0&\mathbb{Z}\to \mathbb{Z}&0\to 0&0\to\mathbb{Z}\\
&&&&n\mapsto 2n&&&&n\mapsto n&&\\
\hline 
{\rm CI}&S_{-+},\eta_{-+}&{\rm L_r \to \rm P}&0\to \mathbb{Z}&0\to 0&2\mathbb{Z}\to \mathbb{Z}&0\to 0&\mathbb{Z}_2\to \mathbb{Z}&\mathbb{Z}_2\to 0&\mathbb{Z}\to \mathbb{Z}&0\to 0\\
&&&&&n\mapsto 2n&&&&n\mapsto n&\\
&&{\rm L_i \to \rm P}&2\mathbb{Z}\to \mathbb{Z}&0\to 0&\mathbb{Z}_2\to \mathbb{Z}&\mathbb{Z}_2\to 0&\mathbb{Z}\to \mathbb{Z}&0\to 0&0\to\mathbb{Z}&0\to 0\\
&&&n\mapsto 2n&&&&n\mapsto n&&&\\
\hline \hline
\end{array}
$$
}
\end{sidewaystable}

\begin{sidewaystable}
\caption*{(Continued)}
\centering
{\scriptsize
$$
\begin{array}{ccccccccccccc}
\mbox{AZ class} & \mbox{Add. symm.}& \mbox{Gap} &d=0&d=1&d=2&d=3&d=4&d=5&d=6&d=7 \\
\hline \hline
{\rm A}&\eta&{\rm L_r} \to {\rm P}&\mathbb{Z}\oplus\mathbb{Z}\to\mathbb{Z}&0\to 0&\mathbb{Z}\oplus\mathbb{Z}\to\mathbb{Z}&0\to 0&\mathbb{Z}\oplus\mathbb{Z}\to\mathbb{Z}&0\to 0&\mathbb{Z}\oplus\mathbb{Z}\to\mathbb{Z}&0\to 0\\
&&&(n,m)\mapsto n-m&&(n,m)\mapsto n-m&&(n,m)\mapsto n-m&&(n,m)\mapsto n-m&\\ 
&& {\rm L_i} \to {\rm P}&0 \to \mathbb{Z}&\mathbb{Z} \to 0&0 \to \mathbb{Z}&\mathbb{Z} \to 0&0 \to \mathbb{Z}&\mathbb{Z} \to 0&0 \to \mathbb{Z}&\mathbb{Z} \to 0\\
&&&&&&&&&&&\\ 
\hline
{\rm AIII}&S_+,\eta_{+}&{\rm L_r,L_i}\to {\rm P}&0\to 0&\mathbb{Z}\oplus\mathbb{Z}\to\mathbb{Z}&0\to 0&\mathbb{Z}\oplus\mathbb{Z}\to\mathbb{Z}&0\to 0&\mathbb{Z}\oplus\mathbb{Z}\to\mathbb{Z}&0\to 0&\mathbb{Z}\oplus\mathbb{Z}\to\mathbb{Z}\\ 
&&&&(n,m)\mapsto n-m&&(n,m)\mapsto n-m&&(n,m)\mapsto n-m&&(n,m)\mapsto n-m\\ 
\hline
{\rm AI}&\eta_-&{\rm L_r} \to {\rm P}&\mathbb{Z}\to\mathbb{Z}_2&0\to\mathbb{Z}_2&\mathbb{Z}\to\mathbb{Z}&0\to 0&\mathbb{Z}\to 0&0\to 0&\mathbb{Z}\to 2\mathbb{Z}&0\to 0\\ 
&&&n\mapsto n&&n\mapsto 2n&&&&n\mapsto n&\\ 
&& {\rm L_i} \to {\rm P}&0 \to \mathbb{Z}_2&\mathbb{Z}_2 \to \mathbb{Z}_2&\mathbb{Z}_2 \to \mathbb{Z}&\mathbb{Z} \to 0&0\to 0&0\to 0&0\to 2\mathbb{Z}&2\mathbb{Z}\to 0\\ 
&&&&n \mapsto 0&&&&&&\\
\hline
{\rm BDI}&S_{--},\eta_{--}&{\rm L_r}\to {\rm P}&0\to 0&\mathbb{Z}\to\mathbb{Z}_2&0\to\mathbb{Z}_2&\mathbb{Z}\to\mathbb{Z}&0\to 0&\mathbb{Z}\to 0&0\to 0&\mathbb{Z}\to 2\mathbb{Z}\\ 
&&&&n\mapsto n&&n\mapsto 2n&&&&n\mapsto n\\ 
&& {\rm L_i}\to {\rm P}&0\to 0&\mathbb{Z}_2\oplus\mathbb{Z}_2\to \mathbb{Z}_2&\mathbb{Z}_2\oplus \mathbb{Z}_2\to\mathbb{Z}_2&\mathbb{Z}\oplus \mathbb{Z}\to\mathbb{Z}&0\to 0&0\to 0&0\to 0&2\mathbb{Z}\oplus 2\mathbb{Z}\to 2\mathbb{Z}\\ 
&&&&(n,m)\mapsto n+m&(n,m)\mapsto n+m&(n,m)\mapsto n+m&&&&(n,m)\mapsto n+m\\ 
\hline
{\rm D}&\eta_-& {\rm L_r} \to {\rm P}&\mathbb{Z}\to 2\mathbb{Z}&0\to 0&\mathbb{Z}\to\mathbb{Z}_2&0\to\mathbb{Z}_2&\mathbb{Z}\to\mathbb{Z}&0\to 0&\mathbb{Z}\to 0&0\to 0\\ 
&&&n\mapsto n&&n\mapsto n&&n\mapsto 2n&&\\
&&{\rm L_i} \to {\rm P}&0\to 2\mathbb{Z}&\mathbb{Z}_2 \to 0&\mathbb{Z}_2 \to \mathbb{Z}_2&\mathbb{Z} \to \mathbb{Z}_2&0 \to \mathbb{Z}&0 \to 0&0 \to 0&2\mathbb{Z}\to 0\\ 
&&&&&n \mapsto n&n \mapsto n&&&&\\ 
\hline
{\rm DIII}&S_{++},\eta_{--}& {\rm L_r,L_i}\to {\rm P}&0\to 0&\mathbb{Z}\to 2\mathbb{Z}&0\to 0&\mathbb{Z}\to\mathbb{Z}_2&0\to\mathbb{Z}_2&\mathbb{Z}\to\mathbb{Z}&0\to 0&\mathbb{Z}\to 0\\ 
&&&&n\mapsto n&&n\mapsto n&&n\mapsto 2n&&\\ 
\hline
{\rm AII}&\eta_-& {\rm L_r} \to {\rm P}&\mathbb{Z}\to 0&0\to 0&\mathbb{Z}\to 2\mathbb{Z}&0\to 0&\mathbb{Z}\to\mathbb{Z}_2&0\to\mathbb{Z}_2&\mathbb{Z}\to\mathbb{Z}&0\to 0\\ 
&&&&&n\mapsto n&&n\mapsto n&&n\mapsto 2n&\\
&& {\rm L_i} \to {\rm P}&0 \to 0&0 \to 0&0 \to 2\mathbb{Z}&2\mathbb{Z} \to 0&0 \to \mathbb{Z}_2&\mathbb{Z}_2 \to \mathbb{Z}_2&\mathbb{Z}_2 \to \mathbb{Z}&\mathbb{Z} \to 0\\ 
&&&&&&&&n \mapsto 0&&\\
\hline 
{\rm CII}&S_{--},\eta_{--}& {\rm L_r} \to {\rm P}&0\to 0&\mathbb{Z}\to 0&0\to 0&\mathbb{Z}\to 2\mathbb{Z}&0\to 0&\mathbb{Z}\to\mathbb{Z}_2&0\to\mathbb{Z}_2&\mathbb{Z}\to\mathbb{Z}\\ 
&&&&&&n\mapsto n&&n\mapsto n&&n\mapsto 2n\\ 
&& {\rm L_i} \to {\rm P}&0\to 0&0\to 0&0\to 0&2\mathbb{Z}\oplus 2\mathbb{Z}\to 2\mathbb{Z}&0\to 0&\mathbb{Z}_2\oplus\mathbb{Z}_2\to \mathbb{Z}_2&\mathbb{Z}_2\oplus \mathbb{Z}_2\to\mathbb{Z}_2&\mathbb{Z}\oplus \mathbb{Z}\to\mathbb{Z}\\ 
&&&&&&(n,m)\mapsto n+m&&(n,m)\mapsto n+m&(n,m)\mapsto n+m&(n,m)\mapsto n+m\\ 
\hline
{\rm C}&\eta_-& {\rm L_r} \to {\rm P}&\mathbb{Z}\to \mathbb{Z}&0\to 0&\mathbb{Z}\to 0&0\to 0&\mathbb{Z}\to 2\mathbb{Z}&0\to 0&\mathbb{Z}\to\mathbb{Z}_2&0\to\mathbb{Z}_2\\ 
&&&n\mapsto 2n&&&&n\mapsto n&&n\mapsto n\\
&& {\rm L_i} \to {\rm P}&0 \to \mathbb{Z}&0 \to 0&0 \to 0&2\mathbb{Z} \to 0&0\to 2\mathbb{Z}&\mathbb{Z}_2 \to 0&\mathbb{Z}_2 \to \mathbb{Z}_2&\mathbb{Z} \to \mathbb{Z}_2\\ 
&&&&&&&&&n \mapsto n&n \mapsto n\\ 
\hline 
{\rm CI}&S_{++},\eta_{--}& {\rm L_r,L_i} \to {\rm P}&0\to \mathbb{Z}_2&\mathbb{Z}\to \mathbb{Z}&0\to 0&\mathbb{Z}\to 0&0\to 0&\mathbb{Z}\to 2\mathbb{Z}&0\to 0&\mathbb{Z}\to\mathbb{Z}_2\\ 
&&&&n\mapsto 2n&&&&n\mapsto n&&n\mapsto n\\ 
\hline \hline
\end{array}
$$
}
\end{sidewaystable}

\begin{sidewaystable}
\caption*{(Continued)}
\centering
{\scriptsize
$$
\begin{array}{ccccccccccccc}
\mbox{AZ class} & \mbox{Add. symm.}& \mbox{Gap} &d=0&d=1&d=2&d=3&d=4&d=5&d=6&d=7 \\
\hline \hline
{\rm A}&S&{\rm L} \to {\rm P}&0\to 0&\mathbb{Z}\to\mathbb{Z}\oplus\mathbb{Z}&0\to 0&\mathbb{Z}\to\mathbb{Z}\oplus\mathbb{Z}&0\to 0&\mathbb{Z}\to\mathbb{Z}\oplus\mathbb{Z}&0\to 0&\mathbb{Z}\to\mathbb{Z}\oplus\mathbb{Z}\\ 
&&&&n\mapsto(n,n)&&n\mapsto(n,n)&&n\mapsto(n,n)&&n\mapsto(n,n)\\ 
\hline
{\rm AIII}&S_-,\eta_-&{\rm L_r\to P}&\mathbb{Z}\to\mathbb{Z}\oplus\mathbb{Z}&0\to 0&\mathbb{Z}\to\mathbb{Z}\oplus\mathbb{Z}&0\to 0&\mathbb{Z}\to\mathbb{Z}\oplus\mathbb{Z}&0\to 0&\mathbb{Z}\to\mathbb{Z}\oplus\mathbb{Z}&0\to 0\\
&&&n\mapsto(n,n)&&n\mapsto(n,n)&&n\mapsto(n,n)&&n\mapsto(n,n)&\\
&&{\rm L_i\to P}&\mathbb{Z}\to\mathbb{Z}\oplus\mathbb{Z}&0\to 0&\mathbb{Z}\to\mathbb{Z}\oplus\mathbb{Z}&0\to 0&\mathbb{Z}\to\mathbb{Z}\oplus\mathbb{Z}&0\to 0&\mathbb{Z}\to\mathbb{Z}\oplus\mathbb{Z}&0\to 0\\
&&&n\mapsto(n,-n)&&n\mapsto(n,-n)&&n\mapsto(n,-n)&&n\mapsto(n,-n)&\\
\hline
{\rm AI}&S_+&{\rm L_r \to \rm P}&\mathbb{Z}_2\to\mathbb{Z}_2\oplus\mathbb{Z}_2&\mathbb{Z}\to\mathbb{Z}\oplus\mathbb{Z}&0\to 0&0\to 0&0\to 0&\mathbb{Z}\to\mathbb{Z}\oplus\mathbb{Z}&0\to 0&\mathbb{Z}_2\to\mathbb{Z}_2\oplus\mathbb{Z}_2\\
&&&n\mapsto(n,n)&n\mapsto(n,n)&&&&n\mapsto(n,n)&&n\mapsto(n,n)\\
&&{\rm L_i \to \rm P}&\mathbb{Z}_2\to\mathbb{Z}_2\oplus\mathbb{Z}_2&\mathbb{Z}\to\mathbb{Z}\oplus\mathbb{Z}&0\to 0&0\to 0&0\to 0&\mathbb{Z}\to\mathbb{Z}\oplus\mathbb{Z}&0\to 0&\mathbb{Z}_2\to\mathbb{Z}_2\oplus\mathbb{Z}_2\\
&&&n\mapsto(n,n)&n\mapsto(n,n)&&&&n\mapsto(n,n)&&n\mapsto(n,n)\\
\hline
{\rm BDI}&S_{+-},\eta_{-+}&{\rm L_r \to \rm P}&\mathbb{Z}_2\to\mathbb{Z}_2\oplus\mathbb{Z}_2&\mathbb{Z}_2\to\mathbb{Z}_2\oplus\mathbb{Z}_2&\mathbb{Z}\to\mathbb{Z}\oplus\mathbb{Z}&0\to 0&0\to 0&0\to 0&\mathbb{Z}\to\mathbb{Z}\oplus\mathbb{Z}&0\to 0\\
&&&n\mapsto(n,n)&n\mapsto(n,n)&n\mapsto(n,n)&&&&n\mapsto(n,n)&\\
&&{\rm L_i \to \rm P}&\mathbb{Z}_2\to\mathbb{Z}_2\oplus\mathbb{Z}_2&\mathbb{Z}_2\to\mathbb{Z}_2\oplus\mathbb{Z}_2&\mathbb{Z}\to\mathbb{Z}\oplus\mathbb{Z}&0\to 0&0\to 0&0\to 0&\mathbb{Z}\to\mathbb{Z}\oplus\mathbb{Z}&0\to 0\\
&&&n\mapsto(n,n)&n\mapsto(n,n)&n\mapsto(n,-n)&&&&n\mapsto(n,-n)&\\
\hline
{\rm D}&S_-&{\rm L \to \rm P}&0\to 0&\mathbb{Z}_2\to\mathbb{Z}_2\oplus\mathbb{Z}_2&\mathbb{Z}_2\to\mathbb{Z}_2\oplus\mathbb{Z}_2&\mathbb{Z}\to\mathbb{Z}\oplus\mathbb{Z}&0\to 0&0\to 0&0\to 0&\mathbb{Z}\to\mathbb{Z}\oplus\mathbb{Z}\\
&&&&n\mapsto(n,n)&n\mapsto(n,n)&n\mapsto(n,n)&&&&n\mapsto(n,n)\\
\hline
{\rm DIII}&S_{+-},\eta_{+-}&{\rm L_r \to \rm P}&\mathbb{Z}\to\mathbb{Z}\oplus\mathbb{Z}&0\to 0&\mathbb{Z}_2\to\mathbb{Z}_2\oplus\mathbb{Z}_2&\mathbb{Z}_2\to\mathbb{Z}_2\oplus\mathbb{Z}_2&\mathbb{Z}\to\mathbb{Z}\oplus\mathbb{Z}&0\to 0&0\to 0&0\to 0\\
&&&n\mapsto(n,n)&&n\mapsto(n,n)&n\mapsto(n,n)&n\mapsto(n,n)&&&\\
&&{\rm L_i \to \rm P}&\mathbb{Z}\to\mathbb{Z}\oplus\mathbb{Z}&0\to 0&\mathbb{Z}_2\to\mathbb{Z}_2\oplus\mathbb{Z}_2&\mathbb{Z}_2\to\mathbb{Z}_2\oplus\mathbb{Z}_2&\mathbb{Z}\to\mathbb{Z}\oplus\mathbb{Z}&0\to 0&0\to 0&0\to 0\\
&&&n\mapsto(n,-n)&&n\mapsto(n,n)&n\mapsto(n,n)&n\mapsto(n,-n)&&&\\
\hline
{\rm AII}&S_+&{\rm L_r\to \rm P}&0\to 0&\mathbb{Z}\to\mathbb{Z}\oplus\mathbb{Z}&0\to 0&\mathbb{Z}_2\to\mathbb{Z}_2\oplus\mathbb{Z}_2&\mathbb{Z}_2\to\mathbb{Z}_2\oplus\mathbb{Z}_2&\mathbb{Z}\to\mathbb{Z}\oplus\mathbb{Z}&0\to 0&0\to 0\\
&&&&n\mapsto(n,n)&&n\mapsto(n,n)&n\mapsto(n,n)&n\mapsto(n,n)&&\\
&&{\rm L_i \to \rm P}&0\to 0&\mathbb{Z}\to\mathbb{Z}\oplus\mathbb{Z}&0\to 0&\mathbb{Z}_2\to\mathbb{Z}_2\oplus\mathbb{Z}_2&\mathbb{Z}_2\to\mathbb{Z}_2\oplus\mathbb{Z}_2&\mathbb{Z}\to\mathbb{Z}\oplus\mathbb{Z}&0\to 0&0\to 0\\
&&&&n\mapsto(n,n)&&n\mapsto(n,n)&n\mapsto(n,n)&n\mapsto(n,n)&&\\
\hline
{\rm CII}&S_{+-},\eta{-+}&{\rm L_r \to \rm P}&0\to 0&0\to 0&\mathbb{Z}\to\mathbb{Z}\oplus\mathbb{Z}&0\to 0&\mathbb{Z}_2\to\mathbb{Z}_2\oplus\mathbb{Z}_2&\mathbb{Z}_2\to\mathbb{Z}_2\oplus\mathbb{Z}_2&\mathbb{Z}\to\mathbb{Z}\oplus\mathbb{Z}&0\to 0\\
&&&&&n\mapsto(n,n)&&n\mapsto(n,n)&n\mapsto(n,n)&n\mapsto(n,n)&\\
&&{\rm L_i \to \rm P}&0\to 0&0\to 0&\mathbb{Z}\to\mathbb{Z}\oplus\mathbb{Z}&0\to 0&\mathbb{Z}_2\to\mathbb{Z}_2\oplus\mathbb{Z}_2&\mathbb{Z}_2\to\mathbb{Z}_2\oplus\mathbb{Z}_2&\mathbb{Z}\to\mathbb{Z}\oplus\mathbb{Z}&0\to 0\\
&&&&&n\mapsto(n,-n)&&n\mapsto(n,n)&n\mapsto(n,n)&n\mapsto(n,-n)&\\
\hline
{\rm C}&S_-&{\rm L \to \rm P}&0\to 0&0\to 0&0\to 0&\mathbb{Z}\to\mathbb{Z}\oplus\mathbb{Z}&0\to 0&\mathbb{Z}_2\to\mathbb{Z}_2\oplus\mathbb{Z}_2&\mathbb{Z}_2\to\mathbb{Z}_2\oplus\mathbb{Z}_2&\mathbb{Z}\to\mathbb{Z}\oplus\mathbb{Z}\\
&&&&&&n\mapsto(n,n)&&n\mapsto(n,n)&n\mapsto(n,n)&n\mapsto(n,n)\\
\hline
{\rm CI}&S_{+-},\eta_{+-}&{\rm L_r \to \rm P}&\mathbb{Z}\to\mathbb{Z}\oplus\mathbb{Z}&0\to 0&0\to 0&0\to 0&\mathbb{Z}\to\mathbb{Z}\oplus\mathbb{Z}&0\to 0&\mathbb{Z}_2\to\mathbb{Z}_2\oplus\mathbb{Z}_2&\mathbb{Z}_2\to\mathbb{Z}_2\oplus\mathbb{Z}_2\\
&&&n\mapsto(n,n)&&&&n\mapsto(n,n)&&n\mapsto(n,n)&n\mapsto(n,n)\\
&&{\rm L_i \to \rm P}&\mathbb{Z}\to\mathbb{Z}\oplus\mathbb{Z}&0\to 0&0\to 0&0\to 0&\mathbb{Z}\to\mathbb{Z}\oplus\mathbb{Z}&0\to 0&\mathbb{Z}_2\to\mathbb{Z}_2\oplus\mathbb{Z}_2&\mathbb{Z}_2\to\mathbb{Z}_2\oplus\mathbb{Z}_2\\
&&&n\mapsto(n,-n)&&&&n\mapsto(n,-n)&&n\mapsto(n,n)&n\mapsto(n,n)\\
\hline \hline
\end{array}
$$
}
\end{sidewaystable}

\clearpage
\begin{table}[]
\caption{Classification table of intrinsic point-gap topology for AZ class.}
\label{tab:SFH_quotient_AZ}
\centering
$$
\begin{array}{ccccccccccccc}
\mbox{~AZ class~}&~d=0~&~d=1~&~d=2~&~d=3~&~d=4~&~d=5~&~d=6~&~d=7~ \\
\hline \hline
{\rm A}&0&\mathbb{Z}&0&\mathbb{Z}&0&\mathbb{Z}&0&\mathbb{Z}\\ 
{\rm AIII}&0&0&0&0&0&0&0&0\\
{\rm AI}&0&\mathbb{Z}&0&0&0&2\mathbb{Z}&0&0\\
{\rm BDI}&0&0&0&0&0&0&0&0\\
{\rm D}&0&0&0&\mathbb{Z}&0&0&0&2\mathbb{Z}\\
{\rm DIII}&0&0&0&0&\mathbb{Z}_2&0&0&0\\
{\rm AII}&0&2\mathbb{Z}&0&0&0&\mathbb{Z}&0&0\\
{\rm CII}&0&0&0&0&0&0&0&0\\
{\rm C}&0&0&0&2\mathbb{Z}&0&0&0&\mathbb{Z}\\
{\rm CI}&\mathbb{Z}_2&0&0&0&0&0&0&0\\
\hline
\hline
\end{array}
$$
\end{table}

\begin{table}[]
\caption{Classification table of intrinsic point-gap topology for AZ$^\dag$ class.}
\label{tab:SFH_quotient_AZdag}
\centering
$$
\begin{array}{ccccccccccccc}
\mbox{~AZ$^\dag$ class~}&~d=0~&~d=1~&~d=2~&~d=3~&~d=4~&~d=5~&~d=6~&~d=7~ \\
\hline \hline 
%%%%%%%%%%%%%%%%%%%%%%%%%%%%%%%%%%%%%%%
{\rm AI^\dag}&0&0&0&2\mathbb{Z}&0&\mathbb{Z}_2&\mathbb{Z}_2&\mathbb{Z}\\
{\rm BDI^\dag}&0&0&0&0&0&0&0&0\\ 
{\rm D^\dag}&0&\mathbb{Z}&0&0&0&2\mathbb{Z}&0&0\\
{\rm DIII^\dag}&0&\mathbb{Z}_2&\mathbb{Z}_2&0&0&0&0&0\\ 
{\rm AII^\dag}&0&\mathbb{Z}_2&\mathbb{Z}_2&\mathbb{Z}&0&0&0&2\mathbb{Z}\\
{\rm CII^\dag}&0&0&0&0&0&0&0&0\\ 
{\rm C^\dag}&0&2\mathbb{Z}&0&0&0&\mathbb{Z}&0&0\\
{\rm CI^\dag}&0&0&0&0&0&\mathbb{Z}_2&\mathbb{Z}_2&0\\
\hline
\hline
\end{array}
$$
\end{table}

\begin{table}[]
\caption{Classification table of intrinsic point-gap topology for AZ class with sublattice symmetry or pseudo-Hermiticity. }
\label{tab:SFH_quotient_AZ_add}
\centering
$$
\begin{array}{ccccccccccccc}
\mbox{~AZ class~}&\mbox{~Add. symm.~}&~d=0~&~d=1~&~d=2~&~d=3~&~d=4~&~d=5~&~d=6~&~d=7~ \\
\hline \hline
{\rm A}&\eta&0&0&0&0&0&0&0&0\\
{\rm AIII}&S_+,\eta_+&0&0&0&0&0&0&0&0\\
\hline 
{\rm A}&S&0&\mathbb{Z}&0&\mathbb{Z}&0&\mathbb{Z}&0&\mathbb{Z}\\ 
{\rm AIII}&S_-,\eta_-&\mathbb{Z}_2&0&\mathbb{Z}_2&0&\mathbb{Z}_2&0&\mathbb{Z}_2&0\\
\hline
{\rm AI}&\eta_+&0&0&0&0&0&0&0&0\\ 
{\rm BDI}&S_{++},\eta_{++}&0&0&0&0&0&0&0&0\\ 
{\rm D}&\eta_+&0&0&0&0&0&0&0&0\\ 
{\rm DIII}&S_{--},\eta_{++}&0&0&0&0&0&0&0&0\\ 
{\rm AII}&\eta_+&0&0&0&0&0&0&0&0\\ 
{\rm CII}&S_{++},\eta_{++}&0&0&0&0&0&0&0&0\\ 
{\rm C}&\eta_+&0&0&0&0&0&0&0&0\\ 
{\rm CI}&S_{--},\eta_{++}&0&0&0&0&0&0&0&0\\ 
\hline
{\rm AI}&S_-&0&\mathbb{Z}&0&0&0&\mathbb{Z}&0&0\\
{\rm BDI}&S_{-+},\eta_{+-}&0&0&0&0&\mathbb{Z}_2&0&\mathbb{Z}_2&0\\
{\rm D}&S_+&0&0&0&\mathbb{Z}&0&\mathbb{Z}_2&0&\mathbb{Z}\\
{\rm DIII}&S_{-+},\eta_{-+}&0&0&0&0&\mathbb{Z}_2&0&\mathbb{Z}_2&0\\
{\rm AII}&S_-&0&\mathbb{Z}&0&0&0&\mathbb{Z}&0&0&\\
{\rm CII}&S_{-+},\eta_{+-}&\mathbb{Z}_2&0&\mathbb{Z}_2&0&0&0&0&0\\
{\rm C}&S_+&0&\mathbb{Z}_2&0&\mathbb{Z}&0&0&0&\mathbb{Z}\\
{\rm CI}&S_{-+},\eta_{-+}&\mathbb{Z}_2&0&\mathbb{Z}_2&0&0&0&0&0\\
\hline
{\rm AI}&\eta_-&0&\mathbb{Z}_2&\mathbb{Z}_2&0&0&0&0&0\\
{\rm BDI}&S_{--},\eta_{--}&0&0&0&0&0&0&0&0\\ 
{\rm D}&\eta_-&0&0&0&0&\mathbb{Z}_2&0&0&0\\ 
{\rm DIII}&S_{++},\eta_{--}&0&0&0&0&\mathbb{Z}_2&\mathbb{Z}_2&0&0\\ 
{\rm AII}&\eta_-&0&0&0&0&0&\mathbb{Z}_2&\mathbb{Z}_2&0\\ 
{\rm CII}&S_{--},\eta_{--}&0&0&0&0&0&0&0&0\\ 
{\rm C}&\eta_-&\mathbb{Z}_2&0&0&0&0&0&0&0\\ 
{\rm CI}&S_{++},\eta_{--}&\mathbb{Z}_2&\mathbb{Z}_2&0&0&0&0&0&0\\ 
\hline
{\rm AI}&S_+&\mathbb{Z}_2&\mathbb{Z}&0&0&0&\mathbb{Z}&0&\mathbb{Z}_2\\
{\rm BDI}&S_{+-},\eta_{-+}&\mathbb{Z}_2&\mathbb{Z}_2&\mathbb{Z}_2&0&0&0&\mathbb{Z}_2&0\\
{\rm D}&S_-&0&\mathbb{Z}_2&\mathbb{Z}_2&\mathbb{Z}&0&0&0&\mathbb{Z}\\
{\rm DIII}&S_{+-},\eta_{+-}&\mathbb{Z}_2&0&\mathbb{Z}_2&\mathbb{Z}_2&\mathbb{Z}_2&0&0&0\\
{\rm AII}&S_+&0&\mathbb{Z}&0&\mathbb{Z}_2&\mathbb{Z}_2&\mathbb{Z}&0&0\\
{\rm CII}&S_{+-},\eta{-+}&0&0&\mathbb{Z}_2&0&\mathbb{Z}_2&\mathbb{Z}_2&\mathbb{Z}_2&0\\
{\rm C}&S_-&0&0&0&\mathbb{Z}&0&\mathbb{Z}_2&\mathbb{Z}_2&\mathbb{Z}\\
{\rm CI}&S_{+-},\eta_{+-}&\mathbb{Z}_2&0&0&0&\mathbb{Z}_2&0&\mathbb{Z}_2&\mathbb{Z}_2\\
\hline \hline
\end{array}
$$
\end{table}

%\bibliography{NH-topo-BBC}

\begin{thebibliography}{116}
\bibitem [{con()}]{converse}%
  \BibitemOpen
  \href@noop {} {}\bibinfo {note} {{Intrinsic point-gap topology is the only
  origin of the non-Hermitian skin effect. Suppose a non-Hermitian system with
  open boundaries exhibits the skin effect and its spectrum is different from
  $\sigma \left( H \left( k \right) \right)$. This, together with Eq.~(\ref{eq:
  inclusion OBC-SIBC}), implies the presence of boundary states under the
  semi-infinite boundary condition, which in turn requires nontrivial point-gap
  topology of $H \left( k \right)$ from Theorem~I.}}\BibitemShut {Stop}%
\end{thebibliography}

\end{document}